\documentclass[a4paper,USenglish,cleveref, autoref, thm-restate]{lipics-v2021}
\usepackage{tikz}
\usetikzlibrary{calc}
\usetikzlibrary{patterns}
\usetikzlibrary{decorations.pathreplacing}
\usepackage[linesnumbered,ruled,vlined]{algorithm2e}

\title{An Improved Approximation Algorithm for the Maximum Weight Independent Set Problem in \texorpdfstring{$d$}{d}-Claw Free Graphs} 

\titlerunning{An Improved Approximation Algorithm for the MWIS in \texorpdfstring{$d$}{d}-Claw Free Graphs} 

\author{Meike Neuwohner}{Forschungsinstitut f\"ur Diskrete Mathematik, Universit\"at Bonn}{neuwohner@or.uni-bonn.de}{https://orcid.org/0000-0002-3664-3687}{}

\authorrunning{M. Neuwohner} 

\Copyright{Meike Neuwohner} 

\nolinenumbers



\relatedversion{} 

\supplement{}

\hideLIPIcs  

\EventEditors{Markus Bl\"{a}ser and Benjamin Monmege}
\EventNoEds{2}
\EventLongTitle{38th International Symposium on Theoretical Aspects of Computer Science (STACS 2021)}
\EventShortTitle{STACS 2021}
\EventAcronym{STACS}
\EventYear{2021}
\EventDate{March 16--19, 2021}
\EventLocation{Saarbr\"{u}cken, Germany (Virtual Conference)}
\EventLogo{}
\SeriesVolume{187}
\ArticleNo{11}
\newtheorem{notation}[theorem]{Notation}
        \newcommand*{\chrg}[2]{\mathrm{charge}(#1,#2)}
        \newcommand*{\chrgp}[2]{\mathrm{charge}'(#1,#2)}
        \newcommand*{\contr}[2]{\mathrm{contr}(#1,#2)}
\begin{document}

\maketitle

\begin{abstract}
In this paper, we consider the task of computing an independent set of maximum weight in a given $d$-claw free graph $G=(V,E)$ equipped with a positive weight function $w:V\rightarrow\mathbb{R}^+$. In doing so, $d\geq 2$ is considered a constant. The previously best known approximation algorithm for this problem is the local improvement algorithm \emph{SquareImp} proposed by Berman \cite{Berman}. It achieves a performance ratio of $\frac{d}{2}+\epsilon$ in time $\mathcal{O}(|V(G)|^{d+1}\cdot(|V(G)|+|E(G)|)\cdot (d-1)^2\cdot \left(\frac{d}{2\epsilon}+1\right)^2)$ for any $\epsilon>0$, which has remained unimproved for the last twenty years. By considering a broader class of local improvements, we obtain an approximation ratio of $\frac{d}{2}-\frac{1}{63,700,992}+\epsilon$ for any $\epsilon>0$ at the cost of an additional factor of $\mathcal{O}(|V(G)|^{(d-1)^2})$ in the running time. In particular, our result implies a polynomial time $\frac{d}{2}$-approximation algorithm. Furthermore, the well-known reduction from the weighted $k$-Set Packing Problem to the Maximum Weight Independent Set Problem in $k+1$-claw free graphs provides a $\frac{k+1}{2}-\frac{1}{63,700,992}+\epsilon$-approximation algorithm for the weighted $k$-Set Packing Problem for any $\epsilon>0$. This improves on the previously best known approximation guarantee of $\frac{k+1}{2}+\epsilon$ originating from the result of Berman \cite{Berman}.
\end{abstract}
\newpage
 \section{Introduction}
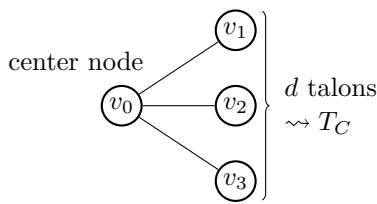
\begin{figure}[t]
  \centering
		\begin{tikzpicture}[scale = 0.5,elem/.style = {circle, draw = black, thick, fill = none, inner sep = 0.5mm, minimum size = 4mm}]
		\node[elem] (C) at (0,0){$v_0$};
		\node[elem] (A1) at (3,2){$v_1$};
		\node[elem] (A2) at (3,0){$v_2$};
		\node[elem] (A3) at (3,-2){$v_3$};
		\draw (C)--(A1);
		\draw (C)--(A2);
		\draw (C)--(A3);
		\draw[ decoration = brace, decorate] (3.7,2.5) to (3.7,-2.5);
		\node at (-1.2,1.2) { center node};
		\node[align = left] at (5.5,0){ $d$ talons\\
		$\rightsquigarrow T_C$};
		\end{tikzpicture}
		\caption{a $d$-claw $C$ for $d=3$}\label{FigClaw}
	\end{figure}

\begin{figure}[t]
 \centering
 \begin{subfigure}{\textwidth}
 \centering
	\begin{tikzpicture}
	\node (11) at (-2,0){};
	\node (1) at (0,2) {$1$};
	\node (2) at (2,2) {$2$};
	\node (3) at (4,2) {$3$};
	\node (4) at (6,2) {$4$};
	\node (5) at (8,2) {$5$};
	\node (13) at (0,0) {$\{1,3\}$};
	\node (23) at (2,0) {$\{2,3\}$};
	\node (33) at (4,0) {$\{3\}$};
	\node (43) at (6,0) {$\{4,3\}$};
	\node (53) at (8,0) {$\{5,3\}$};
	\draw (1)--(13);
	\draw (2)--(23);
	\draw[orange] (3)--(33);
	\draw (4)--(43);
	\draw (5)--(53);
	\draw[orange] (3)--(13);
	\draw[orange] (3)--(23);
	\draw[orange] (3)--(43);
	\draw[orange] (3)--(53);
	\end{tikzpicture}\caption{Example for a claw in the tight instance for $d=6$. It does not improve $A$.}\end{subfigure}
	\begin{subfigure}{\textwidth}
	\centering\begin{tikzpicture}
	\node (1) at (0,2){$1$};
	\node (11) at (-2,0) {$\{1\}$};
	\node (2) at (2,2) {$2$};
	\node (3) at (4,2) {$3$};
	\node (4) at (6,2) {$4$};
	\node (5) at (8,2) {$5$};
	\node (13) at (0,0) {$\{1,3\}$};
	\node (23) at (2,0) {$\{2,3\}$};
	\node (33) at (4,0) {$\{3\}$};
	\node (43) at (6,0) {$\{4,3\}$};
	\node (53) at (8,0) {$\{5,3\}$};
	\draw[green!70!black] (1)--(13);
	\draw (2)--(23);
	\draw[green!70!black] (3)--(33);
	\draw (4)--(43);
	\draw (5)--(53);
	\draw[green!70!black] (3)--(13);
	\draw (3)--(23);
	\draw (3)--(43);
	\draw[green!70!black] (1) --(11);
	\end{tikzpicture}\caption{$\{\{1\},\{1,3\},\{3\}\}$ constitutes a local improvement of constant size.}\end{subfigure}
	\caption{(Part of) the tight instance provided in \cite{Berman}.}\label{FigTightExample}
\end{figure}
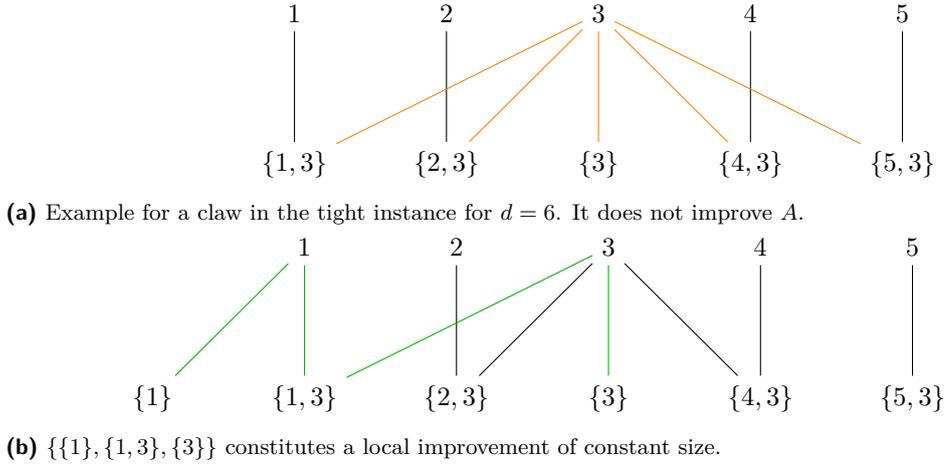

 For $d\geq 1$, a \emph{$d$-claw} $C$ \cite{Berman} is defined to be a star consisting of one center node and a set $T_C$ of $d$ additional vertices connected to it, which are called the \emph{talons} of the claw (see Figure~\ref{FigClaw}). Moreover, similar to \cite{Berman}, we define a $0$-claw to be a graph consisting only of a single vertex $v$, which is regarded as the unique element of $T_C$ in this case. An undirected graph $G=(V,E)$ is said to be \emph{$d$-claw free} if none of its induced subgraphs forms a $d$-claw. For example, $1$-claw free graphs do not possess any edges, while $2$-claw free graphs are disjoint unions of cliques.
 For natural numbers $k\geq 3$, the Maximum Weight Independent Set Problem (MWIS) in $k+1$-claw free graphs is often studied as a generalization of the weighted $k$-Set Packing Problem, which is defined as follows: Given a family $\mathcal{S}$ of sets each of size at most $k$ together with a positive weight function $w:\mathcal{S}\rightarrow\mathbb{R}^+$, the task is to find a disjoint sub-collection of $\mathcal{S}$ of maximum weight. By considering the \emph{conflict graph} $G_\mathcal{S}$ associated with an instance of the weighted $k$-Set Packing Problem, the vertices of which are given by the sets in $\mathcal{S}$ and the edges of which represent non-empty set intersections, one obtains a weight preserving one-to-one correspondence between feasible solutions to the $k$-Set Packing Problem and independent sets in $G_\mathcal{S}$, which can be shown to be $k+1$-claw free.\\
 While as far as the weighted version of the $k$-Set Packing Problem is concerned, the algorithm devised by Berman in 2000 \cite{Berman} to deal with the MWIS in $k+1$-claw free graphs remains unchallenged so far, considerable progress has been made for the cardinality variant during the last decade. The first improvement over the approximation guarantee of $k$ achieved by a simple greedy approach was obtained by Hurkens and Schrijver in 1989 \cite{HurkensSchrijver}, who showed that for any $\epsilon>0$, there exists a constant $p_\epsilon$ for which a local improvement algorithm that first computes a maximal collection of disjoint sets and then repeatedly applies local improvements of constant size at most $p_\epsilon$, until no more exist, yields an approximation guarantee of $\frac{k}{2}+\epsilon$. In this context, a disjoint collection $X$ of sets contained in the complement of the current solution $A$ is considered a \emph{local improvement of size $|X|$} if the sets in $X$ intersect at most $|X|-1$ sets from $A$, which are then replaced by the sets in $X$, increasing the cardinality of the found solution. Hurkens and Schrijver also proved that a performance guarantee of $\frac{k}{2}$ is best possible for a local search algorithm only considering improvements of constant size, while Hazan, Safra and Schwartz \cite{LowerBoundKSetPacking} established in 2006 that no $o(\frac{k}{\log k})$-approximation algorithm is possible in general unless $P=NP$. At the cost of a quasi-polynomial runtime, Halld\'orsson \cite{Halldorsson} could prove an approximation factor of $\frac{k+2}{3}$ by applying local improvements of size logarithmic in the total number of sets. Cygan, Grandoni and Mastrolilli \cite{CyganGrandoniMastrolilli} managed to get down to an approximation factor of $\frac{k+1}{3}+\epsilon$, still with a quasi-polynomial runtime.\\ The first polynomial time algorithm improving on the result by Hurkens and Schrijver was obtained by Sviridenko and Ward \cite{SviridenkoWard} in 2013. By combining means of color coding with the algorithm presented in \cite{Halldorsson}, they achieved an approximation ratio of $\frac{k+2}{3}$. This result was further improved to $\frac{k+1}{3}+\epsilon$ for any fixed $\epsilon>0$ by Cygan \cite{Cygan}, obtaining a polynomial runtime doubly exponential in $\frac{1}{\epsilon}$. The best approximation algorithm for the unweighted $k$-Set Packing Problem in terms of performance ratio and running time is due to F\"urer and Yu from 2014 \cite{FurerYu}, who achieved the same approximation guarantee as Cygan, but a runtime that is only singly exponential in $\frac{1}{\epsilon}$.\\
 Concerning the unweighted version of the MWIS in $d$-claw free graphs, as remarked in \cite{SviridenkoWard}, both the result of Hurkens and Schrijver as well as the quasi-polynomial time algorithms by Halld\'orsson and Cygan, Grandoni and Mastrolilli  translate to this more general context, yielding approximation guarantees of $\frac{d-1}{2}+\epsilon$, $\frac{d+1}{3}$ and $\frac{d}{3}+\epsilon$, respectively. However, it is not clear how to extend the color coding approach relying on coloring the underlying universe to the setting of $d$-claw free graphs \cite{SviridenkoWard}.\\
 When it comes to the weighted variant of the problem, even less is known.  For $d\leq 3$, it is solvable in polynomial time (see \cite{Minty} and \cite{sbihi1980algorithme} for the unweighted, \cite{nakamura2001revision} for the weighted variant), while for $d\geq 4$, again no $o(\frac{d}{\log d})$-approximation algorithm is possible unless $P=NP$ \cite{LowerBoundKSetPacking}. Moreover, in contrast to the unit weight case, considering local improvements the size of which is bounded by a constant can only slightly improve on the performance ratio of $d-1$ obtained by the greedy algorithm since Arkin and Hassin have shown that such an approach yields an approximation ratio no better than $d-2$ in general \cite{ArkinHassin}. Thereby, analogously to the unweighted case, given an independent set $A$, an independent set $X$ is called a \emph{local improvement of $A$} if it is disjoint from $A$ and the total weight of the neighbors of $X$ in $A$ is strictly smaller than the weight of $X$. Despite the negative result in \cite{ArkinHassin}, Chandra and Halld\'orsson \cite{ChandraHalldorsson} have found that if one does not perform the local improvements in an arbitrary order, but in each step augments the current solution $A$ by an improvement $X$ that maximizes the ratio between the total weight of the vertices added to and removed from $A$ (if exists), the resulting algorithm, which the authors call \emph{BestImp}, approximates the optimum solution within a factor of $\frac{2d}{3}$. By scaling and truncating the weight function to ensure a polynomial number of iterations, they obtain a $\frac{2d}{3}+\epsilon$-approximation algorithm for the MWIS in $d$-claw free graphs for any $\epsilon>0$.\\
 As already mentioned, the currently best known approximation guarantee for the MWIS in $d$-claw free graphs is due to Berman \cite{Berman}, who suggested the algorithm \emph{SquareImp}, which iteratively applies local improvements of the squared weight function that arise as sets of talons of claws in $G$, until no more exist. An induced subgraph $C$ of $G$ is thereby called a \emph{claw in $G$} if there is some $t\geq 0$ such that $C$ constitutes a $t$-claw. The algorithm SquareImp achieves an approximation ratio of $\frac{d}{2}$, leading to a polynomial time $\frac{d}{2}+\epsilon$-approximation algorithm for any $\epsilon >0$. Its running time can be bounded by $\mathcal{O}(|V(G)|^{d+1}\cdot(|V(G)|+|E(G)|)\cdot (d-1)^2\cdot \left(\frac{d}{2\epsilon}+1\right)^2)$.\\
   Berman also provides an example for $w\equiv 1$ showing that his analysis is tight. It consists of a bipartite graph $G=(V,E)$ the vertex set of which splits into a maximal independent set $A=\{1,\dots,d-1\}$ such that no claw improves $|A|$, and an optimum solution $B=\binom{A}{1}\cup\binom{A}{2}$, whereby the set of edges is given by $E=\{\{a,b\}:a\in A, b\in B, a\in b\}$. As the example uses unit weights, he also concludes that applying the same type of local improvement algorithm for a different power of the weight function does not provide further improvements.\\ However, as also implied by the result in \cite{HurkensSchrijver}, while no small improvements \emph{forming the set of talons of a claw} in the input graph exist in the tight example given by Berman, once this additional condition is dropped, improvements of small constant size can be found quite easily (see Figure~\ref{FigTightExample}). This in turn indicates that considering a less restricted class of local improvements may result in a better approximation guarantee.\\
In this paper, we revisit the analysis of the algorithm SquareImp proposed by Berman and show that whenever it is close to being tight, the instance actually bears a similar structure to the tight example given in \cite{Berman} in a certain sense. By further observing that if this is the case, there must exist a local improvement (with respect to the squared weight function) of size at most $d-1+(d-1)^2$, we can conclude that a local improvement algorithm looking for improvements of $w^2$ obeying the aforementioned size bound achieves an improved approximation ratio at the cost of an additional $\mathcal{O}(|V(G)|^{(d-1)^2})$ factor in the running time.\\
 The rest of this paper is organized as follows:
 In Section~\ref{SecPrelim}, we review the algorithm SquareImp by Berman and give a short overview of the analysis pointing out the results we reuse in the analysis of our algorithm. The latter is presented in Section~\ref{SecAlgo}, which also provides a detailed analysis proving an approximation guarantee of $\frac{d}{2}-\frac{1}{63,700,992}+\epsilon$ for any $\epsilon>0$. Finally, Section~\ref{SecRemarks} concludes the paper with some remarks on possibilities to improve on the given result, but also difficulties that one might face along the way.
 \section{Preliminaries\label{SecPrelim}}
 In this section, we shortly recap the definitions and main results from \cite{Berman} that we will employ in the analysis of our local improvement algorithm. We first introduce some basic notation that is needed for its formal description.
 \begin{definition}[neighborhood \cite{Berman}]
Given an undirected graph $G=(V,E)$ and subsets $U,W\subseteq V$ of vertices, we define the \emph{neighborhood} $N(U,W)$ of $U$ in $W$ as 
\[N(U,W):=\{w\in W:\exists u\in U: \{u,w\}\in E \vee u=w\}.\]
In order to simplify notation, for $u\in V$ and $W\subseteq V$, we write $N(u,W)$ instead of $N(\{u\}, W)$.\label{DefNeighborhood}
\end{definition}
\begin{notation}
Given a weight function $w:V\rightarrow\mathbb{R}$ and some $U\subseteq V$, we write \linebreak[4]$w^2(U):=\sum_{u\in U} w^2(u)$. Observe that in general, $w^2(U)\neq (w(U))^2$.
\end{notation}
\begin{definition}[\cite{Berman}]
Given an undirected graph $G=(V,E)$, a positive weight function $w:V\rightarrow\mathbb{R}^+$ and an independent set $A\subseteq V$, we say that a vertex set $B\subseteq V$ \emph{improves $w^2(A)$} if $B$ is independent in $G$ and $w^2(A\backslash N(B,A)\cup B) > w^2(A)$ holds.
For a claw $C$ in $G$, we say that $C$ improves $w^2(A)$ if its set of talons $T_C$ does.	
\end{definition}
Observe that an independent set $B$ improves $A$ if and only if we have $w^2(B)>w^2(N(B,A))$ (see Proposition~\ref{PropEquivDefLocalImpr}). Further note that we do not require $B$ to be disjoint from $A$.\\ Using the notation introduced above, Berman's algorithm SquareImp \cite{Berman} can now be formulated as in Algorithm~\ref{AlgoSquareImp}.
\begin{algorithm}[t]
	\DontPrintSemicolon
\KwIn{an undirected $d$-claw free graph $G=(V,E)$ and a positive weight function $w:V\rightarrow\mathbb{R}^+$\;}
\KwOut{an independent set $A\subseteq V$\;}
$A\gets \emptyset$\;
\While{there exists a claw $C$ in $G$ that improves $w^2(A)$}
{$A\gets A\backslash N(T_C, A)\cup T_C$\;}
\Return $A$\;
\caption{SquareImp \cite{Berman}}\label{AlgoSquareImp}	
\end{algorithm}
Observe that by positivity of the weight function, every $v\not\in A$ such that $A\cup\{v\}$ is independent constitutes the talon of a $0$-claw improving $w^2(A)$, so the algorithm returns a maximal independent set.\\
The main idea of the analysis of SquareImp presented in \cite{Berman} is to charge the vertices in $A$ for preventing adjacent vertices in an optimum solution $A^*$ from being included into $A$. The latter is done by spreading the weight of the vertices in $A^*$ among their neighbors in the maximal independent set $A$ in such a way that no vertex in $A$ receives more than $\frac{d}{2}$ times its own weight. The suggested distribution of weights thereby proceeds in two steps:\\ First, each vertex $u\in A^*$ invokes costs of $\frac{w(v)}{2}$ at each $v\in N(u,A)$, leaving a remaining weight of $w(u)-\frac{w(N(u,A))}{2}$ to be distributed. (Note that this term can be negative.)\\ In a second step, each vertex in $u$ therefore sends an amount of $w(u)-\frac{w(N(u,A))}{2}$ to a heaviest neighbor it possesses in $A$, which is captured by the following definition of \emph{charges}:
\begin{definition}[charges \cite{Berman}]
Let $G=(V,E)$ be an undirected graph and let $w:V\rightarrow\mathbb{R}^+$ be a positive weight function. Further assume that an independent set $A^*\subseteq V$ and a maximal independent set $A\subseteq V$ are given. We define a map $\mathrm{charge}:A^*\times A\rightarrow \mathbb{R}$ as follows:\\
For each $u\in A^*$, pick a vertex $v\in N(u,A)$ of maximum weight and call it $n(u)$. Observe that this is possible, because $A$ is a maximal independent set in $G$, implying that $N(u,A)\neq \emptyset$ since either $u\in A$ itself or $u$ possesses a neighbor in $A$.\\
Next, for $u\in A^*$ and $v\in A$, define \[\chrg{u}{v}:=\begin{cases}
w(u)-\frac{1}{2}w(N(u,A)) &,\text{ if }v=n(u)\\
0&,\text{ otherwise }
\end{cases}.\] \label{DefCharges}
\end{definition}
The definition of charges directly implies the subsequent statement:
\begin{corollary}[\cite{Berman}]
 In the situation of Definition~\ref{DefCharges}, we have
 \begin{align*}w(A^*)&=\sum_{u\in A^*} \frac{w(N(u,A))}{2}+\sum_{u\in A^*} \chrg{u}{n(u)}\\&\leq \sum_{u\in A^*} \frac{w(N(u,A))}{2}+\sum_{u\in A^*: \chrg{u}{n(u)}>0} \chrg{u}{n(u)}.\end{align*} \label{CorBoundwAstar}
\end{corollary}

The analysis proposed by Berman now proceeds by bounding the total weight  sent to the vertices in $A$ during the two steps of the cost distribution separately. Lemma~\ref{LemWeightsNeighborhoodsdminus1} thereby bounds the weight received in the first step, while Lemma~\ref{LemPropPositiveCharges} and Lemma~\ref{LemBoundCharges}  take care of the total charges invoked. (Note that although we have slightly changed the formulation of the subsequent results to suit our purposes, they either appear in \cite{Berman} in an equivalent form or are directly implied by the proofs presented there.)
\begin{lemma}[\cite{Berman}]
 In the situation of Definition~\ref{DefCharges}, if the graph $G$ is $d$-claw free for some $d\geq 2$, then
 \[\sum_{u\in A^*} \frac{w(N(u,A))}{2}\leq \frac{d-1}{2}\cdot w(A).\] \label{LemWeightsNeighborhoodsdminus1}
\end{lemma}

\begin{lemma}[\cite{Berman}]
  In the situation of Definition~\ref{DefCharges}, for $u\in A^*$ and $v\in A$ with $\chrg{u}{v}>0$, we have 
 \[w^2(u) - w^2(N(u,A)\backslash\{v\}) \geq 2\cdot\chrg{u}{v}\cdot w(v). \]\label{LemPropPositiveCharges}
\end{lemma}
\begin{lemma}[\cite{Berman}]
Let $G=(V,E)$ be $d$-claw free, $d\geq 2$, and $w:V\rightarrow\mathbb{R}^+$. Let further $A^*$ be an independent set in $G$ of maximum weight and let $A$ be independent in $G$ with the property that no claw improves $w^2(A)$.
Then for each $v\in A$, we have \[\sum_{u\in A^*:\chrg{u}{v}>0}\chrg{u}{v}\leq \frac{w(v)}{2}. \]\label{LemBoundCharges}\end{lemma}
The proofs can be found in the appendix.\\
By combining Corollary~\ref{CorBoundwAstar} with the previous lemmata, one obtains Theorem~\ref{TheoApproxFactor}, stating an approximation guarantee of $\frac{d}{2}$:
\begin{theorem}[\cite{Berman}]
 Let $G=(V,E)$ be $d$-claw free, $d\geq 2$, and $w:V\rightarrow\mathbb{R}^+$. Let further $A^*$ be an independent set in $G$ of maximum weight and let $A$ be independent in $G$ with the property that no claw improves $w^2(A)$.
 Then \[w(A^*)\leq \sum_{u\in A^*} \frac{w(N(u,A))}{2}+\sum_{u\in A^*: \chrg{u}{n(u)}>0} \chrg{u}{n(u)}\leq \frac{d}{2}\cdot w(A). \]\label{TheoApproxFactor}
\end{theorem}

After having recapitulated the results from \cite{Berman} that we will reemploy in our analysis, we are now prepared to study our algorithm that takes into account a broader class of local improvements.
 \section{Improving the Approximation Factor\label{SecAlgo}}
 \subsection{The Local Improvement Algorithm}
 \begin{definition}[Local improvement]
 	Given a $d$-claw free graph $G=(V,E)$, a strictly positive weight function $w:V\rightarrow\mathbb{R}^{+}$ and an independent set $A\subseteq V$, we call an independent set $X\subseteq V$ a \emph{local improvement} of $w^2(A)$ if $|X|\leq (d-1)^2+(d-1)$ and $w^2(A\backslash N(X,A)\cup X) > w^2(A)$.\label{DefLocalImprovement}
 \end{definition}
 \begin{proposition}
  Let $G$, $w$ and $A$ be as in Definition~\ref{DefLocalImprovement}. If $X$ is a local improvement of $w^2(A)$, then $A\backslash N(X,A)\cup X$ is independent in $G$.\label{PropMaintainIndependentSet}
 \end{proposition}
\begin{proposition}
Let $G$, $w$ and $A$ be as in Definition~\ref{DefLocalImprovement}. Then an independent set $X$ of size at most $(d-1)^2+(d-1)$ constitutes a local improvement of $A$ if and only if we have
$w^2(N(X,A))<w^2(X)$.\label{PropEquivDefLocalImpr}
\end{proposition}
\begin{proof}
 By Definition~\ref{DefNeighborhood}, we have $N(X,A)\subseteq A$ and $(A\backslash N(X,A))\cap X=\emptyset$, so 
 \begin{align*}w^2(A\backslash N(X,A)\cup X)&=w^2(A\backslash N(X,A))+w^2(X)\\&=w^2(A)-w^2(N(X,A))+w^2(X),\end{align*} implying the claim.
\end{proof}

 The remainder of Section~\ref{SecAlgo} is now dedicated to the analysis of Algorithm~\ref{LocalImprovementAlgo} for the Maximum Weight Independent Set Problem in $d$-claw free graphs for $d\geq 2$.
 \begin{algorithm}[t]
 		\DontPrintSemicolon
\KwIn{an undirected $d$-claw free graph $G=(V,E)$ and a positive weight function $w:V\rightarrow\mathbb{R}^+$\;}
\KwOut{an independent set $A\subseteq V$\;}
 	$A\gets\emptyset$ \;
 	\While{there exists a local improvement $X$ of $w^2(A)$}{
 	$A\gets A\backslash N(X,A)\cup X$	}
 	\Return A\;
 	\caption{Local improvement algorithm}\label{LocalImprovementAlgo}
 \end{algorithm}
 Thereby, the main result of this paper is given by the following theorem:
 \begin{theorem}
 If $A^*$ is an optimum solution to the MWIS in a $d$-claw free graph $G$ for some $d\geq 2$ and $A$ denotes the solution returned by Algorithm~\ref{LocalImprovementAlgo}, then we have \[w(A^*)\leq \left(\frac{d}{2}-\frac{1}{63,700,992}\right)\cdot w(A).\]
 \end{theorem}
 First, note that Algorithm~\ref{LocalImprovementAlgo} is correct in the sense that it returns an independent set. This follows immediately from the fact that we maintain the property that $A$ is independent throughout the algorithm, because $\emptyset$ is independent and Proposition~\ref{PropMaintainIndependentSet} tells us that none of our update steps can harm this invariant.\\
 Next, observe that Algorithm~\ref{LocalImprovementAlgo} is guaranteed to terminate since no set $A$ can be attained twice, given that $w^2(A)$ strictly increases in each iteration of the while-loop, and there are only finitely many possibilities. Furthermore, each iteration runs in polynomial (considering $d$ a constant) time $\mathcal{O}(|V|^{(d-1)^2+d-1}\cdot (|V|+|E|))$, because there are only $\mathcal{O}(|V|^{(d-1)^2+d-1})$ many possible choices for $X$ and we can check in linear time $\mathcal{O}(|V|+|E|)$ whether a given one constitutes a local improvement.\\ In order to achieve a polynomial number of iterations, we scale and truncate the weight function as explained in \cite{ChandraHalldorsson} and \cite{Berman}. Given a constant $N>1$, we first  compute a greedy solution $A'$ and rescale the weight function $w$ such that $w(A')=N\cdot |V|$ holds. Then, we delete vertices $v$ of truncated weight $\lfloor w(v) \rfloor = 0$ and run Algorithm~\ref{LocalImprovementAlgo} with the integral weight function $\lfloor w\rfloor$. In doing so, we know that $\lfloor w\rfloor^2(A)$ equals zero initially and must increase by at least one in each iteration. On the other hand, at each point, we have \[\lfloor w\rfloor ^2(A)\leq w^2(A)\leq (w(A))^2\leq (d-1)^2w^2(A')=(d-1)^2\cdot N^2\cdot |V|^2, \]which bounds the total number of iterations by the latter term. Finally, if $r>1$ specifies the approximation guarantee achieved by Algorithm~\ref{LocalImprovementAlgo}, $A$ denotes the solution it returns and $A^*$ is an independent set of maximum weight with respect to the original respectively the scaled, but untruncated weight function $w$, we know that \[r\cdot w(A)\geq r\cdot\lfloor w\rfloor (A)\geq\lfloor w\rfloor(A^*)\geq w(A^*)-|A^*|\geq w(A^*)-|V|\geq  \frac{N-1}{N}\cdot w(A^*), \]so the approximation ratio increases by a factor of at most $\frac{N}{N-1}$.
 \subsection{Analysis of the Performance Ratio}
 We now move on to the analysis of the approximation guarantee. Denote some optimum solution by $A^*$ and denote the solution found by Algorithm~\ref{LocalImprovementAlgo} by $A$. Observe that by positivity of the weight function, $A$ must be a maximal independent set, as adding a vertex would certainly yield a local improvement of $w^2(A)$.\\
We first show that for $d= 2$, our algorithm is actually optimal, so that we can restrict ourselves to the case $d\geq 3$ for the main analysis. As already remarked earlier, $2$-claw free graphs are disjoint unions of cliques, so an optimum solution can be found by picking a vertex of maximum weight from each clique. But this is precisely what Algorithm~\ref{LocalImprovementAlgo} does:\\ First, we know that it returns a maximal independent set $A$, which must hence contain exactly one vertex per clique.\\ Second, if for some of the cliques, $A$ contains a vertex $v$ the weight of which is not maximum among all vertices in the clique, and $u\not\in A$ belongs to the same clique and has maximum weight, then $\{u\}$ constitutes a local improvement of $w^2$ since we have $N(u,A)=\{v\}$ and $w^2(v)<w^2(u)$. This contradicts the termination criterion of our algorithm.
 Hence, Algorithm~\ref{LocalImprovementAlgo} is optimum for $d= 2$, and we can assume $d\geq 3$ in the following.\\
For the analysis, we define two constants, $\delta$ and $\epsilon$, which we choose to be $\delta:=\frac{1}{6}$ and $\epsilon:=\frac{1}{5308416}$. These choices satisfy a bunch of inequalities that are used throughout the analysis and can be found in Appendix~\ref{SecAppendixConstants}.\\
Our goal is to show that Algorithm~\ref{LocalImprovementAlgo} produces a $\frac{d-\epsilon\delta}{2}$-approximation.
 We use some notation as well as most of the analysis of the algorithm SquareImp by Berman.
 In particular, we employ the same definition of neighborhoods and charges. Observe that this is well-defined as we have seen that the solution $A$ returned by our algorithm must constitute a maximal independent set in the given graph.\\ For the remainder of this section, fix $d\geq 3$ and some instance of the MWIS in $d$-claw free graphs given by a ($d$-claw free) graph $G=(V,E)$ and a positive weight function \linebreak[4]$w:V\rightarrow\mathbb{R}^+$ and pick an optimum solution $A^*$ for the given instance. Let further $A$ denote the solution returned by Algorithm~\ref{LocalImprovementAlgo}. We have to prove that $w(A^*)\leq \frac{d-\epsilon\delta}{2}\cdot w(A)$.
In doing so, the first step of the analysis is to ensure that for almost all vertices $u\in A^*$, the total weight of their neighborhood in $A$ is only by a small constant factor larger than the weight of $u$. For this purpose, we consider the set $P$ of \grqq payback vertices\grqq\space  $u\in A^*$ for which the total weight of $N(u,A)$ is at least three times as large as $w(u)$. For these vertices, the first step of the weight distribution employed in the analysis by Berman significantly overestimates their weight in that they invoke total costs that are by a factor of $1.5$ larger. As a consequence, we can reduce the total weight sent to $A$ by at least $\frac{w(P)}{2}$, making each of the vertices in $P$ \grqq pay back\grqq\space the unnecessary costs they have created, and still obtain an upper bound on $w(A^*)$. But this means that the analysis of Berman, applied to our algorithm, can actually only be close to tight if the total weight of $P$ is almost zero, which is the essential statement of the following lemma.
 \begin{lemma}
  Let $P:=\{u\in A^*: w(N(u,A))\geq 3\cdot w(u)\}$. Then for all $\gamma>0$, if $w(P)\geq \gamma\cdot w(A)$, we have $w(A^*)\leq \frac{d-\gamma}{2}\cdot w(A)$.\label{LemSizeP}
 \end{lemma}
In order to prove an approximation factor of $\frac{d-\epsilon\delta}{2}$, we can hence restrict ourselves to the case where $w(P)< \epsilon\delta\cdot w(A)$ in the following.\\
Our next goal is to examine the structure of the neighborhoods $N(v,A^*)$ of vertices $v\in A$ that receive a total amount of charges that is close to $\frac{w(v)}{2}$, that is, for which the analysis of SquareImp, applied to Algorithm~\ref{LocalImprovementAlgo}, is almost tight. More precisely, we only consider those neighbors of $v$ sending positive charges to $v$ and try to relate them to the vertices of the form $\{i\}$ respectively $\{i,j\}$ for $i\neq j$ (which actually invoke zero charges in the given instance) from the tight example. For this purpose, the following definitions are required:
\begin{definition}[$T_v$]
 For $v\in A$, we define $T_v:=\{u\in A^*: \chrg{u}{v}>0\}.$
\end{definition}

 \begin{definition}[single vertex]
  For $v\in A$, we call a vertex $u\in T_v$ \emph{single} if 
  \begin{romanenumerate}
   \item $\frac{w(u)}{w(v)} \in [1-\sqrt{\epsilon}, 1+\sqrt{\epsilon}]$ and 
  \item $w(N(u,A))\leq (1+\sqrt{\epsilon})\cdot w(v)$.
  \end{romanenumerate}

 \end{definition}
 \begin{definition}[double vertex]
  For $v\in A$, we call a vertex $u\in T_v$ \emph{double} if $|N(u,A)|\geq 2$ and for $v_1=v$ and $v_2$ a vertex of maximum weight in $N(u,A)\backslash\{v_1\}$, the following properties hold:
  \begin{romanenumerate}
   \item $\frac{w(u)}{w(v_1)}\in[1-\sqrt{\epsilon},1+\sqrt{\epsilon}]$
   \item $\frac{w(v_2)}{w(v_1)}\in [1-\sqrt{\epsilon}, 1]$ and
   \item $(2-\sqrt{\epsilon})\cdot w(v_1)\leq w(N(u,A))<2\cdot w(u)$.
  \end{romanenumerate}

 \end{definition}
Note that for $v_1$ and $v_2$ as in the previous definition, we have \mbox{$w(v_2)\leq w(v_1)$} since we know that $v_1=v=n(u)$ is an element of $N(u,A)$ of maximum weight by definition of $T_v$ and charges.
Further observe that no vertex can be both single and double since this would imply $(2-\sqrt{\epsilon})\cdot w(v)\leq w(N(u,A))\leq (1+\sqrt{\epsilon})\cdot w(v)$ and therefore $2-\sqrt{\epsilon}\leq 1+\sqrt{\epsilon}$, as $w(v)>0$, leading to $\epsilon\geq \frac{1}{4}$ contradicting \eqref{constants5}.\\ The single vertices can be thought of as the vertices of the form $\{i\}$ from the tight example, while the double vertices are in correspondence with those vertices given by sets of size $2$, although in the given example, these actually would not be considered double themselves since they send zero charges.
  \begin{lemma}
 	 For $v\in A$, we either have $\sum_{u\in T_v} \chrg{u}{v} \leq \frac{1-\epsilon}{2}\cdot w(v)$, or for each $u\in T_v$, we have exactly one of the following:
 	\begin{romanenumerate}
 		\item $u$ is single or
 		\item $u$ is double,
 	\end{romanenumerate} 
 	and moreover, there exists at most one $u\in T_v$ that is single.\label{NeighbourhoodLargeCharges}	
 \end{lemma}
  We would like to provide some motivation why we are actually interested in a statement of this type.
To this end, first note that if the total weight of those vertices $v\in A$ satisfying $\sum_{u\in T_v} \chrg{u}{v} \leq \frac{1-\epsilon}{2}\cdot w(v)$ constitutes some constant fraction of $w(A)$, we get an improved approximation factor since we gain an $\frac{\epsilon}{2}$-fraction of the weight of each such vertex when bounding the weight of $A^*$. On the other hand, if there are only few such vertices (in terms of weight), the vertices $v\in A$ for which the analysis of SquareImp is almost tight when it comes to charges, and for which all vertices in the set $T_v$ can hence be classified as being either single or double, possess a large total weight. The set comprising these vertices $v$ can be further split into the collection of those vertices that feature a neighbor that is single, and the set of those who do not. In order to gain some intuitive understanding of why Algorithm~\ref{LocalImprovementAlgo} achieves a better approximation guarantee than SquareImp, we have to see how both types of vertices can be helpful for our analysis.\\ For this purpose, let us first consider those vertices $v\in A$ all neighbors (in $T_v$) of which are double. Observe that for a double vertex $u_0\in A^*$, its neighborhood $N(u_0,A)$ consists of two vertices $v_1=n(u_0)$ and $v_2$ of roughly the same weight as $u_0$, plus maybe some additional vertices the total weight of which is by a factor in the order of $\sqrt{\epsilon}$ smaller. For simplicity, imagine that $v_1$ and $v_2$ have exactly the same weight and that there are no further neighbors of $u_0$ in $A$. In this situation, it is completely arbitrary whether $v_1$ or $v_2$ is chosen as $n(u_0)$. In particular, we can bound both of the terms $w^2(u_0)-w^2(N(u_0,A)\backslash\{v_1\})$ and $w^2(u_0)-w^2(N(u_0,A)\backslash\{v_2\})$ by $2\cdot\chrg{u_0}{n(u_0)}\cdot w(v_1)=2\cdot\chrg{u_0}{n(u_0)}\cdot w(v_2)$ from below. Moreover, the proof of Lemma~\ref{LemBoundCharges} tells us that for each $v\in A$, we actually get the stronger statement \[\sum_{u\in N(v,A^*)}\max\{0,w^2(u)-w^2(N(u,A)\backslash\{v\})\}\leq w^2(v). \]When summing over all $v\in A$, while every vertex $u\in A^*$ adds at least $2\cdot\chrg{u}{n(u)}$ by Lemma~\ref{LemPropPositiveCharges}, our \grqq ideal\grqq\space double vertex $u_0$ actually contributes twice as much since it adds an amount of at least $2\cdot\chrg{u}{n(u)}\cdot w(v_{1/2})$ for both $v_1$ and $v_2$.\\ Although for general double vertices, the situation is more complicated, one can still show that $w^2(u)-w^2(N(u,A)\backslash\{v_1\})$ amounts to almost $3\cdot\chrg{u}{v_1}\cdot w(v_1)$, or $u$ adds approximately $\chrg{u}{v_1}\cdot w(v_2)$ when it comes to $v_2$. As a consequence, for those vertices $v\in A$ receiving a total amount of charges of at least $\frac{1-\epsilon}{2}\cdot w(v)$ and all neighbors of which are double, the total charges sent to $v$ can be counted almost three instead of only two times, resulting in an improved approximation factor provided the total weight of these vertices constitutes a constant fraction of $w(A)$.\\We are therefore left with discussing the role of those $v\in A$ that possess at least one single neighbor. By Lemma~\ref{NeighbourhoodLargeCharges}, we further know that those $v$ have exactly one single neighbor, which we denote by $t(v)$ in the following. Recall that by definition of single vertices, this neighbor bears roughly the same weight as $v$, and $v$ makes up almost all of $N(t(v),A)$ in terms of weight. Imagine removing each such vertex $v$ with a single neighbor from $A$ and its neighbor $t(v)\in T_v$ from $A^*$. Then the sets of vertices removed from $A$ and $A^*$, respectively, have roughly the same weight. It further constitutes a large fraction of $w(A)$, provided that $w(P)$, as well as the total weight of vertices for which the analysis of SquareImp is not close to being tight and the total weight of vertices with only double neighbors are small. (Remember that we obtain a better approximation guarantee if this is not the case.) But now, given that the ratio between the weights of the sets of 
vertices we have removed from $A$ and $A^*$, respectively, is close to $1$, we 
must get an improved approximation guarantee unless the ratio between the weights of the sets of vertices $A'^*$ and $A'$ remaining from $A^*$ and $A$ is way larger than $\frac{d}{2}$. But then, we know that we can find a local improvement $X$ of $w^2(A')$ in the resulting instance, which can be extended to a local improvement in the original one by adding vertices that were removed from $A^*$ to make up for the additional weight of neighbors of $X$ that were removed from $A$. The existence of this local improvement contradicts the termination criterion of Algorithm~\ref{LocalImprovementAlgo}.\\
 We have therefore outlined the key ideas of the analysis of Algorithm~\ref{LocalImprovementAlgo} and in particular convinced ourselves of the benefit of the lemma. Its proof can be found in the appendix.
After having seen that all neighbors of vertices $v$ for which the analysis of SquareImp, applied to our algorithm, is almost tight, are either double or single, we continue by establishing the \grqq usefulness\grqq\space of double vertices. As already outlined before, we show that the charges invoked by these can be counted almost three instead of only two times, which is captured by the next lemma.
\begin{lemma}
 Let $u\in T_v$ be double, let $v=v_1$ and let $v_2$ be a vertex of maximum weight in $N(u,A)\backslash\{v_1\}$. Then at least one of the following inequalities holds:
 \begin{romanenumerate}
  \item $w^2(u)-w^2(N(u,A)\backslash\{v_1\})\geq \frac{149}{50}\cdot\chrg{u}{v_1}\cdot w(v_1)$ or
  \item $w^2(u)-w^2(N(u,A)\backslash\{v_2\})\geq \frac{49}{50}\cdot\chrg{u}{v_1}\cdot w(v_2)$.
 \end{romanenumerate}
\label{LemRatioCharges}
\end{lemma}

\noindent When motivating Lemma~\ref{NeighbourhoodLargeCharges}, we proposed to add charges invoked by vertices in $A^*$ to a certain extent for vertices in $A$. This rather vague idea is clarified by the next definition as well as the two propositions and the lemma it is followed by.\\
While Proposition~\ref{PropUpperBoundContr} bounds the total amount the neighborhood of each $v\in A$ can contribute to $v$ in a locally optimal solution, Proposition~\ref{PropContrCharge} and Lemma~\ref{LemContrDouble} give lower bounds on the fraction of the invoked charges non-double and double vertices contribute in total.
\begin{definition}[contribution]
 Define a contribution map \\$\mathrm{contr}:A^*\times A\rightarrow\mathbb{R}_{\geq 0}$ by setting
 \[\mathrm{contr}(u,v):=\begin{cases}\max\left\{0,\frac{w^2(u)-w^2(N(u,A)\backslash\{v\})}{w(v)}\right\} &, \text{ if $v\in N(u,A)$}\\
                         0 &,\text{ else}
                        \end{cases}.\]
\end{definition}
\begin{proposition}
 For each $v\in A$, we have $\sum_{u\in A^*}\mathrm{contr}(u,v)\leq w(v)$.\label{PropUpperBoundContr}
\end{proposition}
\begin{proposition}
 For each $u\in A^*$, we have \[\sum_{v\in A}\contr{u}{v}\geq \contr{u}{n(u)}\geq 2\cdot\chrg{u}{n(u)}.\]\label{PropContrCharge}
\end{proposition}
\begin{lemma}
 For each double vertex $u$, we have $\sum_{v\in A}\mathrm{contr}(u,v)\geq \frac{149}{50}\cdot\chrg{u}{n(u)}.$\label{LemContrDouble}
\end{lemma}
 \begin{definition}[$C$ and $D$]
Let $C$ denote the set of all $v\in A$ for which \begin{romanenumerate} \item $\sum_{u\in T_v}\chrg{u}{v}>\frac{1-\epsilon}{2}\cdot w(v)$ and \item all vertices in $T_v$ are double.\end{romanenumerate}
Let further $D:=\bigcup_{v\in C} T_v$.
\end{definition}
Note that all vertices in $D$ are double by definition. The following proposition tells us that the total charges invoked by vertices in $D$ constitute a considerable fraction of the weight of $C$.
\begin{proposition}
 $\sum_{u\in D}\chrg{u}{n(u)}\geq \frac{1-\epsilon}{2}\cdot w(C).$\label{PropChargesWC}
\end{proposition}
As we have seen that double vertices contribute a factor of at least $\frac{149}{50}$ times the charges they send, we can finally conclude that we obtain an improved approximation factor unless the weight of $C$ is extremely small compared to $w(A)$, which is the statement of the next lemma.
\begin{lemma}
 If $w(C)\geq \frac{25}{12}\cdot\epsilon\delta\cdot w(A)$, then $w(A^*)\leq\frac{d-\epsilon\delta}{2}\cdot w(A)$.\label{LemWeightOfCSmall}
\end{lemma}
By the previous lemma, we know that we can assume $w(C)<\frac{25}{12}\cdot\epsilon\delta\cdot w(A)$ in the following.
As outlined before, we continue by proving that we get the desired approximation guarantee if the set of vertices for which the analysis of SquareImp is not almost tight constitutes at least a $\delta$ fraction of the weight of $A$.
 Let therefore \[\bar{B}:= \left\{v\in A:\sum_{u\in T_v}\chrg{u}{v}>\frac{1-\epsilon}{2}\cdot w(v)\right\} \]denote the set of vertices for which the analysis of SquareImp is close to being tight.
 \begin{lemma}
 If $w(\bar{B})\leq (1-\delta) \cdot w(A)$, then $\frac{d-\epsilon\delta}{2}\cdot w(A)\geq  w(A^*)$.\label{LemCloseToTightForAlmostAllVertices}	
 \end{lemma}
  If we have $w(\bar{B})\leq (1-\delta) \cdot w(A)$, we achieve the claimed approximation factor of $\frac{d-\epsilon\delta}{2}$, so assume $w(\bar{B})> (1-\delta) \cdot w(A)$ in the following. Let further $B:=\bar{B}\backslash C$. Then we have $w(B)=w(\bar{B})-w(C)>(1-\delta-\frac{25}{12}\cdot\epsilon\delta)\cdot w(A)$. By Lemma~\ref{NeighbourhoodLargeCharges}, each vertex $v\in B$ has a unique neighbor in $T_v$ which is single. Call this neighbor $t(v)$ and let $B^*:=\{t(v), v\in B\}$.
We proceed by proving two lemmata that will later help us to transform local improvements in the instance arising by deleting the vertices in $B$, $B^*$ and $P$ into local improvements in the original one.
 Lemma~\ref{SmallRemainingNeighbourhood} thereby tells us that for each $v\in B$, the total weight of the neighbors of $t(v)$ in $A$ other than $v$ is extremely small, while Lemma~\ref{LemBoundWsqV} establishes a relation between the squared weights of $v$ and $t(v)$.
 \begin{lemma}
 	For $v\in B$, we have $w(N(t(v),A)\backslash\{v\})\leq \sqrt{\epsilon}\cdot w(v)$.\label{SmallRemainingNeighbourhood}
 \end{lemma}
  \begin{lemma}
   For $v\in B$, we have $w(v)^2\leq w(t(v))^2 +(4\sqrt{\epsilon}+4\epsilon)\cdot w^2(v)$. \label{LemBoundWsqV}
  \end{lemma}
 Consider the sets $A':=A\backslash B$ and $A'^*:=A^*\backslash (B^*\cup P)$ that arise from deleting all vertices in $B$ and $B^*\cup P$.
 As outlined before, we would like to apply the analysis of SquareImp to bound the weight of $A'^*$ in terms of the weight of $A'$. However, in order to employ the definition of charges, we have to make sure that $A'$ constitutes a maximal independent set in $G[A'\cup A'^*]$. Showing this property is the purpose of the following lemma.
 \begin{lemma}
 	If there exists a vertex $u\in A'^*$ such that $N(u,A')=\emptyset$, then there exist a local improvement of $w^2(A)$ in the original instance.
 	\label{Amaximal}
 \end{lemma}
Due to the termination criterion of our algorithm, we know that there is no local improvement in the original instance, so the previous lemma tells us that every vertex in $A'^*$ must possess a neighbor in $A'$ (considering vertices as adjacent to themselves), showing that $A'$ is a maximal independent set in $G[A'\cup A'^*]$. We can hence apply the same strategy as in the analysis of SquareImp to bound the weight of $A'^*$ by the weight of $A'$, letting each vertex send charges to its heaviest neighbor in $A'$, which must exist by the previous arguments. More precisely, we apply the definition of charges, Definition~\ref{DefCharges}, to the sub-instance induced by $A'\cup A'^*$, in which $A'^*$ is independent and $A'$ is a maximal independent set. Call the resulting charge map $\mathrm{charge}'$ and recall that it is constructed as follows:\\ For each $u\in A'^*$, we pick a heaviest neighbor $v\in N(u,A')$ and call it $n'(u)$. Then, for $u\in A'^*$ and $v\in A'$, we define \[\chrgp{u}{v}:=\begin{cases} w(u)-\frac{w(N(u,A'))}{2} &\text{ if $v=n'(u)$}\\ 0 &\text{ otherwise}\end{cases}. \]For $v\in A'$, let $T'_v:=\{u\in A'^*: \chrgp{u}{v} >0\}$ denote the set of vertices in $A'^*$ that now send positive charges to $v$.\\
 We show that we obtain the desired approximation ratio, provided \[\sum_{u\in T'_v}\chrgp{u}{v}\leq \frac{d+2}{4}\cdot w(v) \]holds for all $v\in A'$, and that we can find a local improvement of $w^2(A)$ in the original instance if this is not the case, contradicting the fact that our algorithm did terminate.
 \begin{lemma}
  If $\sum_{u\in T'_v}\chrgp{u}{v}\leq \frac{d+2}{4}\cdot w(v)$ holds for all $v\in A'$, then we have $w(A^*)\leq \frac{d-\epsilon\delta}{2}\cdot w(A)$.\label{LemImprovedApproxIfSmallCharges}
 \end{lemma}
We are left with proving the following lemma:
 \begin{lemma}
 For all $v\in A'$, we have \[\sum_{u\in T'_{v}}\chrgp{u}{v}\leq \frac{d+2}{4}\cdot w(v). \]\label{LemGetImprovementOldInstance}
 \end{lemma}
This concludes the proof that Algorithm~\ref{LocalImprovementAlgo} achieves approximation factor of at most \[\frac{d-\epsilon\delta}{2}=\dfrac{d-\frac{1}{31850496}}{2}=\frac{d}{2}-\frac{1}{63700992}. \] By scaling and truncating the weight function , we obtain a polynomial time $\frac{d}{2}-\frac{1}{63700992}+\epsilon'$-approximation algorithm for any $\epsilon'>0$, whereby the running time depends polynomially on $\frac{1}{\epsilon'}$. In particular, setting $\epsilon':=\frac{1}{63700992}$, we get a polynomial time $\frac{d}{2}$-approximation algorithm. However, given the fact that the running time of (at least a straightforward implementation of) Algorithm~\ref{LocalImprovementAlgo} is in $\Omega(|V|^{(d-1)^2+(d-1)})$, this result remains of only theoretical interest for the time being.
 \section{Further Remarks\label{SecRemarks}}
The proven result indicates that an approximation ratio of $\frac{d}{2}$ is not the end of the story of local improvement algorithms for the Maximum Weight Independent Set Problem in $d$-claw free graphs. This observation is inevitably followed by the question of how far one can still get with this approach. Concerning algorithms that only consider local improvements of some fixed constant size (possibly dependent on $d$), the result of Hurkens and Schrijver \cite{HurkensSchrijver} implies a lower bound of $\frac{d-1}{2}$ for $d \geq 4$. This raises the question of whether and how the gap between our result, providing an approximation guarantee of $\frac{d}{2}-\frac{1}{63700992}+\epsilon'$ for any $\epsilon'>0$, and the lower bound of $\frac{d-1}{2}$ can be closed. Although the choice of our constants $\epsilon$ and $\delta$ still permits some room for optimization, as the rather rough estimates in the proof of the properties \eqref{constants1} to \eqref{constants11} indicate, the more critical ones among them still seem to be \grqq tight enough\grqq\space to limit hope for an improvement in an entirely different order of magnitude. Therefore, we also picked our constants in a way keeping the proof of \eqref{constants1}-\eqref{constants11} as short as possible. Some further ideas might be required to get substantially closer to an approximation factor of $\frac{d-1}{2}$. Whether or not the latter is possible could be regarded as a worthwhile subject for further research.
\newpage
\bibliography{set_packing_lipics}
\newpage
\appendix
\section{Proofs of Lemmata from the Analysis of SquareImp}
 \begin{proof}[Proof of Lemma~\ref{LemWeightsNeighborhoodsdminus1}]
 As $A^*$ is independent in $G$, each $v\in V$ satisfies $|N(v,A^*)|\leq d-1$, because either $v\in A^*$ and $N(v,A^*)=\{v\}$, or $v\not\in A^*$ and $N(v,A^*)$ constitutes the set of talons of a claw centered at $v$, provided it is non-empty.
\end{proof}
\begin{proof}[Proof of Lemma~\ref{LemPropPositiveCharges}]
$\chrg{u}{v}>0$ implies $v=n(u)\in N(u,A)$ and, therefore, \begin{align*} w^2(N(u,A)\backslash\{v\})&=\sum_{x\in N(u,A)\backslash\{v\}}w^2(x)\\&\leq \sum_{x\in N(u,A)\backslash\{v\}} w(x)\cdot \max_{y\in N(u,A)} w(y)\\&=w(N(u,A)\backslash\{v\})\cdot w(v)\\&=(w(N(u,A))-w(v))\cdot w(v).\end{align*} From this, we get
 \begin{align*} 2\cdot\chrg{u}{v}\cdot w(v) &= (2\cdot w(u)-w(N(u,A)))\cdot w(v)\\ &= 2\cdot w(u)\cdot w(v)-w(N(u,A))\cdot w(v)\\ &\leq w^2(u)+w^2(v)-w(N(u,A))\cdot w(v)\\ &=w^2(u)-(w(N(u,A))-w(v))\cdot w(v)\\&\leq w^2(u)-w^2(N(u,A)\backslash\{v\})\end{align*} as claimed.
\end{proof}
\begin{proof}[Proof of Lemma~\ref{LemBoundCharges}] Assume for a contradiction that \[\sum_{u\in A^*: \chrg{u}{v}>0} \chrg{u}{v}>\frac{w(v)}{2} \] for some $v\in A$. Then $v\not\in A^*$ since \[\{u\in A^*:\chrg{u}{v}>0\}=\{v\}=N(v,A)=N(v,A^*) \] and \[\sum_{u\in A^*: \chrg{u}{v}>0}\chrg{u}{v}=\chrg{v}{v}=\frac{w(v)}{2} \] otherwise. Hence, $T:=\{u\in A^*: \chrg{u}{v}>0\}$ forms the set of talons of a claw centered at $v$. By Lemma~\ref{LemPropPositiveCharges}, it satisfies  \[w^2(T)=\sum_{u\in T} w^2(u)>\sum_{u\in T} w^2(N(u,A)\backslash \{v\}) +w^2(v)\geq w^2(N(T,A)), \] contradicting the fact that no claw improves $w^2(A)$.\end{proof}
 \section{Inequalities Satisfied by Our Choice of \texorpdfstring{$\epsilon$}{epsilon} and 
\texorpdfstring{$\delta$}{delta}}\label{SecAppendixConstants}
 \begin{equation}
 4-2\cdot\frac{6-9\sqrt{\epsilon}}{4-10\sqrt{\epsilon}}-9\sqrt{\epsilon}\geq\frac{49}{50}\label{constants1}
\end{equation}
 \begin{equation}
 9\cdot(4\sqrt{\epsilon}+5\epsilon)<1 \label{constants2}
 \end{equation}
 \begin{equation}
(1+\sqrt{\epsilon})\cdot\left(1-\delta-\frac{25}{12}\cdot\epsilon\delta\right)+\frac{3d}{4}\cdot \left(\delta+\frac{25}{12}\cdot\epsilon\delta\right) +\epsilon\delta\leq\frac{d-\epsilon\delta}{2} \label{constants3}
 \end{equation}
 \begin{equation}
 36\sqrt{\epsilon}+45\epsilon\leq \frac{1}{32} \label{constants4}
 \end{equation}
 \begin{equation}
 0<\epsilon < \frac{16}{100}<\frac{1}{4} \label{constants5}
 \end{equation}
 \begin{equation}
 1-3\sqrt{\epsilon}>\frac{1}{2} \label{constants6}
 \end{equation}
 \begin{equation}
  1+\sqrt{\epsilon} < \frac{3d}{4}\label{constants7}
 \end{equation}
 \begin{equation}
  4\cdot \left(1-\frac{3}{2}\cdot\sqrt{\epsilon}\right)\cdot(1-\sqrt{\epsilon})\geq 3>\frac{149}{50}\label{constants8}
 \end{equation}
 \begin{equation}
  \frac{49\cdot(1-\epsilon)}{100}\geq \frac{12}{25}\label{constants9}
 \end{equation}

\begin{equation}
(2-10\sqrt{\epsilon})\cdot\frac{6-9\sqrt{\epsilon}}{4-10\sqrt{\epsilon}}\geq\frac{149}{50}
\label{constants10}
\end{equation}
\begin{equation}
\min\{2-10\sqrt{\epsilon},6-9\sqrt{\epsilon},4-10\sqrt{\epsilon}\}=2-10\sqrt{\epsilon}>0\label{constants11}
\end{equation}
\begin{proof}
 \eqref{constants1}:
 \begin{align*}4-2\cdot\frac{6-9\sqrt{\epsilon}}{4-10\sqrt{\epsilon}}-9\sqrt{\epsilon}&\geq 4-2\cdot\dfrac{6}{4-\frac{10}{1000}}-\frac{9}{1000}=4-\frac{1200}{399}-\frac{9}{1000}\\&=\frac{1,596,000-1,200,000-3,591}{399,000}=\frac{392,409}{399,000}\\&>\frac{391,020}{399,000}=\frac{49}{50}\end{align*}
 \eqref{constants2}:
 \[9\cdot(4\sqrt{\epsilon}+5\epsilon)<9\cdot\left(\frac{4}{1000}+\frac{5}{1000000}\right)<1.\]
 \eqref{constants3}:
 	\begin{alignat}{5}
 	&(1+\sqrt{\epsilon})\cdot\left(1-\delta-\frac{25}{12}\cdot\epsilon\delta\right)+\frac{3d}{4}\cdot \left(\delta+\frac{25}{12}\cdot\epsilon\delta\right) +\epsilon\delta&\quad\leq\quad&\frac{d-\epsilon\delta}{2}&\notag\\
 	\Leftrightarrow\quad&(1+\sqrt{\epsilon})\cdot(1-\delta)+\left(\frac{3\delta}{4}+\frac{3}{4}\cdot\frac{25\epsilon\delta}{12}\right)\cdot d&\quad\leq\quad&\frac{d-\epsilon\delta}{2}&\notag\\
 	&+\epsilon\delta\cdot\left(1-\frac{25}{12}\cdot(1+\sqrt{\epsilon})\right)&&&\notag\\
 	\Leftarrow\quad&(1+\sqrt{\epsilon})\cdot(1-\delta)+\left(\frac{3\delta}{4}+\frac{25\epsilon\delta}{16}\right)\cdot d+\epsilon\delta&\quad\leq\quad&\frac{d-\epsilon\delta}{2}&\notag\\
 	\Leftrightarrow\quad&(1+\sqrt{\epsilon})\cdot(1-\delta)+\left(\frac{3\delta}{4}+\frac{25\epsilon\delta}{16}\right)\cdot d+ \frac{3}{2}\cdot\epsilon\delta&\quad\leq\quad&\frac{d}{2}&\quad\text{$|$ $\delta=\frac{1}{6}$}\notag\\
 	\Leftrightarrow\quad&(1+\sqrt{\epsilon})\cdot\frac{5}{6}+\left(\frac{1}{8}+\frac{25\epsilon}{96}\right)\cdot d+\frac{\epsilon}{4}&\quad\leq\quad&\frac{d}{2}&\notag\\
 	\Leftrightarrow\quad&(1+\sqrt{\epsilon})\cdot\frac{5}{6}+\frac{12+25\epsilon}{96}\cdot d+\frac{\epsilon}{4}&\quad\leq\quad&\frac{48}{96}\cdot d&\notag\\
 	\Leftrightarrow\quad&(1+\sqrt{\epsilon})\cdot\frac{5}{6}+\frac{\epsilon}{4}&\quad\leq\quad&\frac{36-25\epsilon}{96}\cdot d&\notag\end{alignat}
 	As $d\geq 3$ and $\epsilon<1$, the latter is implied by 
 	\begin{alignat}{5}
 	&(1+\sqrt{\epsilon})\cdot\frac{5}{6}+\frac{\epsilon}{4}&\quad\leq\quad&\frac{108-75\epsilon}{96}&\notag\\
 	\Leftrightarrow\quad&(1+\sqrt{\epsilon})\cdot 80+24\epsilon&\quad\leq\quad&108-75\epsilon&\notag\\
 	\Leftrightarrow\quad&80\sqrt{\epsilon}+99\epsilon&\quad\leq\quad&28&\quad\text{$|$ $\epsilon<\frac{1}{1000000}$}\notag\\
 	\Leftarrow\quad&\frac{80}{1000}+\frac{99}{1000000}&\quad\leq\quad&28.&\notag
 	\end{alignat}
\eqref{constants4}:
\[36\sqrt{\epsilon}+45\epsilon =\frac{36}{2304}+\frac{45}{5308416}=\frac{1}{64}+\frac{5}{589824}<\frac{1}{32}\]
\eqref{constants5}: clear\\
\eqref{constants6}:
\[1-3\sqrt{\epsilon}>1-\frac{3}{1000}>\frac{1}{2}\]
\eqref{constants7}:
\[1+\sqrt{\epsilon} < 1+\frac{1}{\sqrt{1000000}}=1+\frac{1}{1000}<\frac{9}{4}\leq \frac{3d}{4}\]
\eqref{constants8}:
\begin{align*}
  4\cdot \left(1-\frac{3}{2}\cdot\sqrt{\epsilon}\right)\cdot(1-\sqrt{\epsilon})&\geq 4\cdot \left(1-\frac{3}{2000}\right)\cdot\left(1-\frac{1}{1000}\right)\\
  &=\frac{4\cdot 1997\cdot 999}{2,000,000}=\frac{1,995,003}{500,000}>3>\frac{149}{50}
 \end{align*}
 \eqref{constants9}:
  \[\frac{49\cdot(1-\epsilon)}{100}=\dfrac{49\cdot\left(1-\frac{1}{5308416}\right)}{100}>\dfrac{49\cdot\left(1-\frac{1}{49}\right)}{100}=\frac{48}{100}= \frac{12}{25}\]
  \eqref{constants10}:
\begin{align*}
(2-10\sqrt{\epsilon})\cdot\frac{6-9\sqrt{\epsilon}}{4-10\sqrt{\epsilon}}&>\left(2-\frac{10}{1000}\right)\cdot\dfrac{6-\frac{10}{1000}}{4}=\frac{199\cdot 599}{40,000}=\frac{119,201}{40,000}\\&>\frac{119,200}{40,000}=\frac{149}{50}
\end{align*}
\eqref{constants11}: Follows directly from $\sqrt{\epsilon}<\frac{1}{1000}$.
\end{proof}
\section{Propositions and Proofs Omitted in the Main Part}
 The following proposition is helpful to bound the sizes of candidate local improvements we consider during the analysis.
 \begin{proposition}
  For any $v\in A$, we have $|N(v, A^*)|\leq d-1$ and for any $u\in A^*$, $|N(u,A)|\leq d-1$.\label{PropSizeBound}
 \end{proposition}
\begin{proof}
 For $v\in A$, if further $v\in A^*$, then $N(v,A^*)=\{v\}$, because $A^*$ is independent, and therefore $|N(v,A^*)|=1<2\leq d-1$ since $d\geq 3$. If $N(v,A^*)$ is empty, we are also done, so assume $v\not\in A^*$ and $N(v,A^*)\neq \emptyset$. Then by independence of $A^*$, $N(v,A^*)$ forms the set of talons of a claw in $G$ centered at $v$. Consequently, $d$-claw freeness of $G$ implies the desired size bound. The second statement can be obtained analogously.
\end{proof}
\begin{proof}[Proof of Lemma~\ref{LemSizeP}]
As for any claw in $G$, its set of talons possesses a size that is not larger than $\max\{1,d-1\}=d-1\leq (d-1)^2+(d-1)$ and is therefore considered as a possible improvement during our algorithm,
 Theorem~\ref{TheoApproxFactor} implies that \[\sum_{u\in A^*} \frac{w(N(u,A))}{2}+\sum_{u\in A^*:\chrg{u}{n(u)}>0}\chrg{u}{n(u)}\leq \frac{d}{2}\cdot w(A). \]\\
 By definition of charges, we have $\chrg{u}{n(u)}=w(u)-\frac{w(N(u,A))}{2}$, so 
 \begin{align*}\frac{d}{2}\cdot w(A)\geq &\sum_{u\in A^*} \frac{w(N(u,A))}{2}+\sum_{u\in A^*:\chrg{u}{n(u)}>0}\chrg{u}{n(u)}\\
 =& \sum_{u\in A^*} \frac{w(N(u,A))}{2}+\max\left\{w(u)-\frac{w(N(u,A))}{2},0\right\}\\
  =&\sum_{u\in A^*} \max\left\{w(u),\frac{w(N(u,A))}{2}\right\}\\
  \geq& \sum_{u\in P} \frac{3}{2}\cdot w(u)+\sum_{u\in A^*\backslash P} w(u)\\
  =& w(A^*)+\frac{w(P)}{2}.
 \end{align*}
Therefore, $w(P)\geq \gamma\cdot w(A)$ implies $w(A^*)\leq \frac{d-\gamma}{2}\cdot w(A)$ as claimed.
\end{proof}
\begin{proof}[Proof of Lemma~\ref{NeighbourhoodLargeCharges}]
 	If $\sum_{u\in T_v} \chrg{u}{v} \leq \frac{1-\epsilon}{2}\cdot w(v)$, we are done, so assume the contrary, i.e.\ \begin{equation}
 	\sum_{u\in T_v} \chrg{u}{v} > \frac{1-\epsilon}{2}\cdot w(v). \label{AlmostNice}
 	\end{equation}
 	 We have $|T_v|\subseteq N(v,A^*)$ by definition, so $|T_v|\leq d-1$ by Proposition~\ref{PropSizeBound}. As Algorithm~\ref{LocalImprovementAlgo} has terminated, $T_v$ does not yield a local improvement of $w^2$ and we know that \[\sum_{u\in T_v} w^2(u)=w^2(T_v)\leq w^2(N(T_v,A))\leq w^2(v) +\sum_{u\in T_v}w^2(N(u,A)\backslash\{v\}), \]and the outer inequality is equivalent to 
 	\begin{equation}\sum_{u\in T_v} w^2(u)-w^2(N(u,A)\backslash\{v\}) \leq w^2(v). \label{NoImprovement}\end{equation}
 	By Lemma~\ref{LemPropPositiveCharges}, we know that if $\chrg{u}{v} >0$ (which is the case for all $u\in T_v$ by definition), we have \begin{equation}w^2(u)-w^2(N(u,A)\backslash\{v\})\geq 2\cdot\chrg{u}{v}\cdot w(v).\label{EqGeqCharge}\end{equation}
As $w(v)>0$, for $u\in T_v$, let $\epsilon_u \geq 0$ such that
 	\begin{equation}
 	 w^2(u)-w^2(N(u,A)\backslash\{v\})= 2\cdot\chrg{u}{v}\cdot w(v)+\epsilon_u\cdot w^2(v).\label{DefEpsU}
 	\end{equation}

 	Then \eqref{AlmostNice} and \eqref{NoImprovement} imply
 	\begin{align*}
 	w^2(v)\geq&\sum_{u\in T_v}w^2(u)-w^2(N(u,A)\backslash\{v\})\\
 	= &\sum_{u\in T_v}2\cdot\chrg{u}{v}\cdot w(v)+\epsilon_u\cdot w(v)^2\\
 	> &\quad2\cdot \frac{1-\epsilon}{2}\cdot w^2(v) + \sum_{u\in T_v}\epsilon_u\cdot w^2(v)\\
 	&=w^2(v)\cdot \left(1-\epsilon + \sum_{u\in T_v}\epsilon_u\right), 
 	\end{align*} and $w(v)>0$ yields \begin{equation}
 	                            \sum_{u\in T_v}\epsilon_u\leq \epsilon. \label{SumLeqEps}
 	                           \end{equation} We now show that for each $u\in T_v$, one of the conditions listed in the lemma applies:\\
 	Pick $u\in T_v$. By definition of charges, we know that $v=n(u)$ is a neighbor of $u$ in $A$ of maximum weight, implying
 	\begin{align}
 	 w^2(N(u,A)\backslash\{v\})&=\sum_{x\in N(u,A)\backslash\{v\}} w^2(x)\notag\\
 	 &\leq \sum_{x\in N(u,A)\backslash\{v\}}w(x)\cdot\max\{0,\max_{y\in N(u,A)\backslash\{v\}}w(y)\}\notag\\
 	 &=(w(N(u,A))-w(v))\cdot\max\{0,\max_{y\in N(u,A)\backslash\{v\}}w(y)\},\label{EqBoundsWsqNeighborhood}
 	\end{align}
 	whereby $\max\emptyset:=-\infty$. By \eqref{DefEpsU}, we therefore obtain
 	
 	\begin{alignat}{3}
 	 & \quad w^2(u)-w^2(N(u,A)\backslash\{v\})\quad&=&\quad 2\cdot\chrg{u}{v}\cdot w(v)+\epsilon_u\cdot w^2(v)\notag\\
 	 \Leftrightarrow &\quad w^2(u)-w^2(N(u,A)\backslash\{v\}) \quad&=&\quad (2\cdot w(u)-w(N(u,A)))\cdot w(v)\notag\\
 	 &&&\quad+\epsilon_u\cdot w^2(v)\notag\\
 	 \Leftrightarrow &\quad w^2(u)+w^2(v)-w^2(N(u,A)\backslash\{v\}) \quad&=&\quad (2\cdot w(u)+w(v)-w(N(u,A)))\cdot w(v)\notag\\
 	 &&&\quad+\epsilon_u\cdot w^2(v)\notag,\end{alignat}
 	 which results in 
 	 \[ (w(u)-w(v))^2-w^2(N(u,A)\backslash\{v\})+(w(N(u,A))-w(v))\cdot w(v) = \epsilon_u\cdot w^2(v). \]

 	Applying \eqref{EqBoundsWsqNeighborhood} yields
 	\begin{equation}(w(u)-w(v))^2+(w(N(u,A))-w(v))\cdot(w(v)-\max\{0,\max_{y\in N(u,A)\backslash\{v\}}w(y)\}) \leq \epsilon_u\cdot w^2(v).\label{EqTogetherAtMostEpsUTimesWsqV}\end{equation}
 As both summands in \eqref{EqTogetherAtMostEpsUTimesWsqV} are nonnegative since real squares are nonnegative, $v\in N(u,A)$ is of maximum weight and $w>0$, \eqref{EqTogetherAtMostEpsUTimesWsqV} in particular implies that both \begin{align}
\epsilon_u\cdot w^2(v)&\geq (w(u)-w(v))^2 \label{EqUandVclose}\text{ and}\\\epsilon_u\cdot w^2(v)&\geq (w(N(u,A))-w(v))\cdot(w(v)-\max\{0,\max_{y\in N(u,A)\backslash\{v\}}w(y)\}).\label{EqOnceorTwiceTheWeight}\end{align}
From \eqref{EqUandVclose}, we can infer that $|w(u)-w(v)|\leq\sqrt{\epsilon_u}\cdot w(v)$, which in turn implies that \begin{align*}w(u)&\leq w(v)+|w(u)-w(v)|\leq (1+\sqrt{\epsilon_u})\cdot w(v)\text{ as well as }\\w(v)&\leq w(u)+|w(v)-w(u)|\leq w(u)+\sqrt{\epsilon_u}\cdot w(v),\end{align*} which yields $(1-\sqrt{\epsilon_u})\cdot w(v)\leq  w(u)$. As a consequence, by \eqref{SumLeqEps}, we obtain \begin{equation} \frac{w(u)}{w(v)}\in[1-\sqrt{\epsilon_u}, 1+\sqrt{\epsilon_u}]\subseteq[1-\sqrt{\epsilon},1+\sqrt{\epsilon}].\label{EqIntervalRatioUV}\end{equation}
In addition to that, \eqref{EqOnceorTwiceTheWeight} tells us that at least one of the two inequalities \begin{align} \sqrt{\epsilon_u}\cdot w(v)&\geq w(v)-\max\{0,\max_{y\in N(u,A)\backslash\{v\}}w(y)\}\label{EqVandV2close} \text{ or}\\ \sqrt{\epsilon_u}\cdot w(v)&\geq w(N(u,A))-w(v)\label{EqVMakesUpAlmostWholeNeighborhood}\end{align} must hold. If \eqref{EqVandV2close} applies, the fact that $\epsilon_u\leq \epsilon<1$ by \eqref{constants5} and \eqref{SumLeqEps}, together with $w(v)>0$, implies that $N(u,A)\backslash\{v\}\neq \emptyset$, so let $v_2\in N(u,A)\backslash\{v\}$ be of maximum weight. Then \begin{align}w(v)-w(v_2)&\leq \sqrt{\epsilon_u}\cdot w(v)\text{ and hence}\notag\\ (1-\sqrt{\epsilon})\cdot w(v)&\leq (1-\sqrt{\epsilon_u})\cdot w(v)\leq w(v_2)\leq w(v) \label{EqScndCondDoubleFullfilled}\end{align} by maximality of $w(v)$ in $N(u,A)$. From this, we also get \[(2-\sqrt{\epsilon})\cdot w(v)\leq w(v)+w(v_2)\leq w(N(u,A))<2\cdot w(u), \]whereby the last inequality follows from the fact that $u$ sends positive charges to $v$. Hence, together with \eqref{EqIntervalRatioUV} and \eqref{EqScndCondDoubleFullfilled}, all conditions for $u$ being double are fulfilled. In case \eqref{EqVMakesUpAlmostWholeNeighborhood} holds true, we get \[w(N(u,A))\leq (1+\sqrt{\epsilon_u})\cdot w(v)\leq (1+\sqrt{\epsilon})\cdot w(v), \]leaving us with a vertex that is single by \eqref{EqIntervalRatioUV}.\\
In order to finally see that there can be at most one vertex $u\in T_v$ which is single, observe that for a single vertex $u$, we have \begin{align*}\chrg{u}{v}&=w(u)-\frac{w(N(u,A))}{2}\geq (1-\sqrt{\epsilon})\cdot w(v)-\frac{1+\sqrt{\epsilon}}{2}\cdot w(v)\\&=\frac{1-3\sqrt{\epsilon}}{2}\cdot w(v).\end{align*}
Hence, the existence of at least two single vertices in $T_v$ and \eqref{constants6} would imply
\[\sum_{u\in T_v}\chrg{u}{v}\geq (1-3\sqrt{\epsilon})\cdot w(v)>\frac{w(v)}{2} \]and \eqref{EqGeqCharge}, combined with the fact that $w(v)>0$, would yield
\[\sum_{u\in T_v} w^2(u)-w^2(N(u,A)\backslash\{v\})\geq \sum_{u\in T_v}2\cdot\chrg{u}{v}\cdot w(v)>w^2(v),\] a contradiction to \eqref{NoImprovement}.
\end{proof}
\begin{proof}[Proof of Lemma~\ref{LemRatioCharges}]
 We distinguish two cases:\\ 
 \textbf{Case $1$:} $w(v_1)\geq w(u)$.
 Then we have \begin{align}0\leq w(N(u,A))-w(v_1)&= 2\cdot (w(u)-\chrg{u}{v_1})-w(v_1)\notag\\&= w(u)-2\cdot\chrg{u}{v_1}+w(u)-w(v_1)\notag\\&\leq w(u)-2\cdot \chrg{u}{v_1}\notag\end{align} and therefore
 \begin{align}w^2(u)-w^2(N(u,A)\backslash\{v_1\})&\geq w^2(u)-(w(N(u,A))-w(v_1))^2\notag\\&\geq w^2(u)-(w(u)-2\cdot \chrg{u}{v_1})^2\notag\\&=w^2(u)-w^2(u)+4\cdot w(u)\cdot \chrg{u}{v_1}\notag\\&\phantom{=}-4\cdot\chrg{u}{v_1}^2\notag\\
 &=4\cdot\chrg{u}{v_1}\cdot (w(u)-\chrg{u}{v_1})\label{EqAlmostFourTimesCharges}.\end{align}
 Given that for a double vertex, we have
 \begin{align*}\chrg{u}{v_1}&=w(u)-\frac{w(N(u,A))}{2}\leq w(u)-\frac{2-\sqrt{\epsilon}}{2}\cdot w(v_1)\\
 	&\leq w(u)-\frac{2-\sqrt{\epsilon}}{2(1+\sqrt{\epsilon})}\cdot w(u)\leq w(u)- \frac{(2-\sqrt{\epsilon})\cdot (1-\sqrt{\epsilon})}{2}\cdot w(u)\\&=w(u)\cdot \frac{2-(2-3\sqrt{\epsilon}+\epsilon)}{2}\leq  \frac{3}{2}\cdot\sqrt{\epsilon}\cdot w(u)\end{align*} since $\frac{1}{1+\sqrt{\epsilon}}=1-\frac{\sqrt{\epsilon}}{1+\sqrt{\epsilon}}\geq 1-\sqrt{\epsilon}$, \eqref{EqAlmostFourTimesCharges} implies
 	\[w^2(u)-w^2(N(u,A)\backslash\{v_1\})\geq 4\cdot \left(1-\frac{3}{2}\cdot\sqrt{\epsilon}\right)\cdot w(u)\cdot \chrg{u}{v_1}.\]
 	Further knowing that $w(u)\geq (1-\sqrt{\epsilon})\cdot w(v_1)$, we finally obtain
 	\begin{align*}w^2(u)-w^2(N(u,A)\backslash\{v_1\})&\geq 4\cdot \left(1-\frac{3}{2}\cdot\sqrt{\epsilon}\right)\cdot(1-\sqrt{\epsilon})\cdot w(v_1)\cdot \chrg{u}{v_1}\\ &\geq \frac{149}{50}\cdot \chrg{u}{v_1}\cdot w(v_1)\end{align*} by \eqref{constants8} as claimed.\\ 
 \textbf{Case $2$:} $w(v_1)< w(u)$.
 In this case, we get \begin{align}w^2(u)-w^2(N(u,A)\backslash\{v_1\})&=\quad w^2(u)-w^2(v_2)-w^2(N(u,A)\backslash\{v_1,v_2\})\notag\\
                       &=\quad w^2(u)-(w(u)-(w(u)-w(v_2)))^2\notag\\&\phantom{=}\quad-w^2(N(u,A)\backslash\{v_1,v_2\})\notag\\
                       &=\quad w^2(u)-w^2(u)+2\cdot w(u)\cdot (w(u)-w(v_2))\notag\\&\phantom{=}\quad-(w(u)-w(v_2))^2-w^2(N(u,A)\backslash\{v_1,v_2\})\notag\\
                       &=\quad 2\cdot w(u)\cdot (w(u)-w(v_2))-(w(u)-w(v_2))^2\notag\\&\phantom{=}\quad-w^2(N(u,A)\backslash\{v_1,v_2\}).\label{EqBoundGain}
                      \end{align}
By definition of double vertices and our case assumption, we have \[w(u)>w(v_1)\geq w(v_2)\geq (1-\sqrt{\epsilon})\cdot w(v_1)\geq \frac{1-\sqrt{\epsilon}}{1+\sqrt{\epsilon}}\cdot w(u)\geq (1-2\sqrt{\epsilon})\cdot w(u) \]and
therefore $0<w(u)-w(v_2)\leq 2\sqrt{\epsilon}\cdot w(u)$ and \begin{equation}
(w(u)-w(v_2))^2\leq 2\sqrt{\epsilon}\cdot w(u)\cdot (w(u)-w(v_2)).\label{EqBoundSquareDiff}\end{equation}
In addition to that, we get \begin{align*}w(N(u,A)\backslash\{v_1,v_2\})&=w(N(u,A))-w(v_1)-w(v_2)\\&<2\cdot w(u)-w(v_1)-w(v_2)\\&\leq 2\cdot (w(u)-w(v_2)),\end{align*}
leading to
\begin{align}w^2(N(u,A)\backslash\{v_1,v_2\})&\leq (w(N(u,A)\backslash\{v_1,v_2\}))^2\notag\\&\leq 2\cdot\left(w(u)-\frac{w(v_1)+w(v_2)}{2}\right)\cdot 2\cdot (w(u)-w(v_2))\notag\\  &\leq 2\cdot\left(w(u)-\frac{w(v_1)+w(v_2)}{2}\right)\cdot 4\sqrt{\epsilon}\cdot w(u)\notag\\ &= 8\sqrt{\epsilon}\cdot w(u)\cdot \left(w(u)-\frac{w(v_1)+w(v_2)}{2}\right)\label{EqBoundWeightWithout}\\&\leq 8\sqrt{\epsilon}\cdot w(u)\cdot (w(u)-w(v_2))\label{EqBoundWeightWithout2}.\end{align}
Combining \eqref{EqBoundGain} with $w(v_1)<w(u)$, \eqref{constants11}, \eqref{EqBoundSquareDiff} and \eqref{EqBoundWeightWithout2} results in
\begin{align}
 w^2(u)-w^2(N(u,A)\backslash\{v_1\})&\geq (2-10\sqrt{\epsilon})\cdot w(u)\cdot (w(u)-w(v_2))\notag\\&\geq (2-10\sqrt{\epsilon})\cdot w(v_1)\cdot (w(u)-w(v_2))\label{EqBoundGain2}.
\end{align} As double vertices send positive charges, we further have \begin{align}0<\chrg{u}{v_1}&=w(u)-\frac{w(N(u,A))}{2}\leq w(u)-\frac{w(v_1)+w(v_2)}{2}\notag\\&\leq w(u)-w(v_2).\label{EqChargesWeightDiff}\end{align} 
Let therefore $\alpha\geq 1$ such that \begin{equation}w(u)-w(v_2)= \alpha\cdot\left(w(u)-\frac{w(v_1)+w(v_2)}{2}\right).\label{EqAlpha1}\end{equation}
                                                                                                                         
Then \begin{align}w(u)-w(v_1)&=2\cdot\left(w(u)-\frac{w(v_1)+w(v_2)}{2}\right)-(w(u)-w(v_2))\notag\\&=(2-\alpha)\cdot\left(w(u)-\frac{w(v_1)+w(v_2)}{2}\right).\label{EqAlpha2}\end{align}
                                                                                                                                                                             
Consequently, \eqref{EqBoundGain2}, \eqref{EqChargesWeightDiff} and \eqref{EqAlpha1} yield
\begin{align*}w^2(u)-w^2(N(u,A)\backslash\{v_1\})&\geq (2-10\sqrt{\epsilon})\cdot w(v_1)\cdot (w(u)-w(v_2))\\&\geq (2-10\sqrt{\epsilon})\cdot\alpha\cdot w(v_1)\cdot \left(w(u)-\frac{w(v_1)+w(v_2)}{2}\right)\\&\geq (2-10\sqrt{\epsilon})\cdot\alpha\cdot w(v_1)\cdot \chrg{u}{v_1}.\end{align*}
If $\alpha \geq \frac{6-9\sqrt{\epsilon}}{4-10\sqrt{\epsilon}}$, whereby numerator and denominator are positive by \eqref{constants11}, then we get $(2-10\sqrt{\epsilon})\cdot\alpha\geq \frac{149}{50}$ by \eqref{constants10} and are therefore done.                                                                                                       
We can hence assume $\alpha < \frac{6-9\sqrt{\epsilon}}{4-10\sqrt{\epsilon}}$ in the following. By similar calculations as before, we get \begin{align}w^2(u)-w^2(N(u,A)\backslash\{v_2\})&=w^2(u)-w^2(v_1)-w^2(N(u,A)\backslash\{v_1,v_2\})\notag\\
                       &=w^2(u)-(w(u)-(w(u)-w(v_1)))^2\notag\\&\phantom{=}-w^2(N(u,A)\backslash\{v_1,v_2\})\notag\\
                       &=w^2(u)-w^2(u)+2\cdot w(u)\cdot (w(u)-w(v_1))\notag\\&\phantom{=}-(w(u)-w(v_1))^2-w^2(N(u,A)\backslash\{v_1,v_2\})\notag\\
                       &=2\cdot w(u)\cdot (w(u)-w(v_1))-(w(u)-w(v_1))^2\notag\\&\phantom{=}-w^2(N(u,A)\backslash\{v_1,v_2\}).\label{EqBoundGain3}
                      \end{align}
By definition of double vertices and our case assumption, we have \[(1-\sqrt{\epsilon})\cdot w(u)\leq \left(1-\frac{\sqrt{\epsilon}}{1+\sqrt{\epsilon}}\right)\cdot w(u)=\frac{w(u)}{1+\sqrt{\epsilon}}\leq w(v_1)< w(u),\] implying $0< w(u)-w(v_1)\leq \sqrt{\epsilon}\cdot w(u)$, as well as $w(v_2)\leq w(v_1)<w(u)$, leading to $0< w(u)-w(v_1)\leq w(u)-\frac{w(v_1)+w(v_2)}{2}.$ We therefore get \begin{equation}(w(u)-w(v_1))^2\leq \sqrt{\epsilon}\cdot w(u)\cdot \left(w(u)-\frac{w(v_1)+w(v_2)}{2}\right).\label{EqSquaredWeightDiff2}\end{equation}
Together with \eqref{EqBoundWeightWithout}, \eqref{EqAlpha2} and \eqref{EqSquaredWeightDiff2}, \eqref{EqBoundGain3} leads to
\begin{align*}
 w^2(u)-w^2(N(u,A)\backslash\{v_2\})&=2\cdot w(u)\cdot (w(u)-w(v_1))-(w(u)-w(v_1))^2\\&\phantom{=}-w^2(N(u,A)\backslash\{v_1,v_2\})\\
 &\geq  2\cdot w(u)\cdot(2-\alpha)\cdot\left(w(u)-\frac{w(v_1)+w(v_2)}{2}\right)\\&\phantom{=}-\sqrt{\epsilon}\cdot w(u)\cdot \left(w(u)-\frac{w(v_1)+w(v_2)}{2}\right)\\&\phantom{=}-8\sqrt{\epsilon}\cdot w(u)\cdot \left(w(u)-\frac{w(v_1)+w(v_2)}{2}\right)\\
 &\geq(4-2\alpha-9\sqrt{\epsilon})\cdot w(u)\cdot\left(w(u)-\frac{w(v_1)+w(v_2)}{2}\right)\\
 &\geq(4-2\alpha-9\sqrt{\epsilon})\cdot w(v_2)\cdot\chrg{u}{v_1}\\
 &\geq \frac{49}{50}\cdot w(v_2)\cdot\chrg{u}{v_1},
\end{align*}
whereby the last two inequalities follow from \eqref{constants1}, \eqref{EqChargesWeightDiff}, $\alpha < \frac{6-9\sqrt{\epsilon}}{4-10\sqrt{\epsilon}}$ and our case assumption. This finishes the proof of the lemma.\end{proof}
 \begin{proof}[Proof of Proposition~\ref{PropUpperBoundContr}]
 If $v\in A^*$, this is true, because we get $N(v,A^*)=N(v,A)=\{v\}$ and $\mathrm{contr}(v,v)=w(v)$ in this case.\\
 If $v\not\in A^*$, the set $T$ of vertices sending positive contributions to $v$ constitutes the set of talons of a claw centered at $v$ and $\sum_{u\in T}\mathrm{contr}(u,v)>w(v)$ would imply that $T$ constitutes a local improvement of $w^2$.
\end{proof}
\begin{proof}[Proof of Proposition~\ref{PropContrCharge}]
 The first inequality follows by nonnegativity of the contribution, which also implies the second inequality in case $\chrg{u}{n(u)}\leq 0$. If $\chrg{u}{n(u)}>0$, Lemma~\ref{LemPropPositiveCharges} provides the desired statement.
\end{proof}
\begin{proof}[Proof of Lemma~\ref{LemContrDouble}]
 By Lemma~\ref{LemPropPositiveCharges}, we know that $\contr{u}{n(u)}\geq 2\cdot\chrg{u}{n(u)}$ since by definition of a double vertex, $u\in T_{n(u)}$ sends positive charges to $n(u)$. By Lemma~\ref{LemRatioCharges}, we further know that for $v_1=n(u)$ and $v_2$ an element of $N(u,A)\backslash\{v_1\}$ of maximum weight, we have \begin{romanenumerate}
 \item$\contr{u}{v_1}\geq \frac{149}{50}\cdot\chrg{u}{v_1}$ and $\contr{u}{v_2}\geq 0$ or
 \item $\contr{u}{v_1}\geq 2\cdot\chrg{u}{v_1}$ and $\contr{u}{v_2}\geq \frac{49}{50}\cdot\chrg{u}{v_1}$,\end{romanenumerate} implying $\contr{u}{v_1}+\contr{u}{v_2}\geq \frac{149}{50}\cdot\chrg{u}{v_1}=\frac{149}{50}\cdot\chrg{u}{n(u)}$ in either case.
 Consequently, nonnegativity of the contribution yields
 \[\sum_{v\in N(u,A)}\contr{u}{v}\geq \contr{u}{v_1}+\contr{u}{v_2}\geq \frac{149}{50}\cdot\chrg{u}{n(u)} \]as claimed.
\end{proof}
\begin{proof}[Proof of Proposition~\ref{PropChargesWC}]
As for $u\in T_v$, we have $v=n(u)$ and $T_v$ and $T_{v'}$ are in particular disjoint for $v\neq v'$, we get
 \[\sum_{u\in D}\chrg{u}{n(u)}=\sum_{v\in C}\sum_{u\in T_v}\chrg{u}{v}\geq \frac{1-\epsilon}{2}\cdot w(C) \]by definition of $C$ and $D$.
\end{proof}
\begin{proof}[Proof of Lemma~\ref{LemWeightOfCSmall}]
By Proposition~\ref{PropUpperBoundContr}, Proposition~\ref{PropContrCharge}, Lemma~\ref{LemContrDouble} and Proposition~\ref{PropChargesWC}, we get \begin{align*}w(A)& \geq \sum_{v\in A}\sum_{u\in A^*}\contr{u}{v}=\sum_{u\in A^*}\sum_{v\in A}\contr{u}{v}\\
&=\sum_{u\in D}\sum_{v\in A}\contr{u}{v}+\sum_{u\in A^*\backslash D}\sum_{v\in A}\contr{u}{v}\\
&\geq\sum_{u\in D}\frac{149}{50}\cdot\chrg{u}{n(u)}+\sum_{u\in A^*\backslash D} 2\cdot\chrg{u}{n(u)}\\
&=\sum_{u\in A^*}2\cdot\chrg{u}{n(u)}+\frac{49}{50}\cdot\sum_{u\in D}\chrg{u}{n(u)}\\
&\geq \sum_{u\in A^*}2\cdot \chrg{u}{n(u)}+\frac{49\cdot(1-\epsilon)}{100}\cdot w(C)\\
&\geq \sum_{u\in A^*}2\cdot \chrg{u}{n(u)}+\frac{12}{25}\cdot w(C)\end{align*} by \eqref{constants9},
so $\sum_{u\in A^*}\chrg{u}{n(u)}\leq \frac{w(A)}{2}-\frac{6}{25}\cdot w(C)$, and $w(C)\geq \frac{25}{12}\cdot\epsilon\delta\cdot w(A)$ yields $\sum_{u\in A^*}\chrg{u}{n(u)}\leq \frac{1-\epsilon\delta}{2}\cdot w(A)$. Applying Corollary~\ref{CorBoundwAstar} and Lemma~\ref{LemWeightsNeighborhoodsdminus1} provides the desired bound \[w(A^*)\leq \frac{d-1}{2}\cdot w(A)+\sum_{u\in A^*}\chrg{u}{n(u)}\leq \frac{d-\epsilon\delta}{2}\cdot w(A).\]
\end{proof}
\begin{proof}[Proof of Lemma~\ref{LemCloseToTightForAlmostAllVertices}]
	By \eqref{NoImprovement} and \eqref{EqGeqCharge} from the proof of Lemma~\ref{NeighbourhoodLargeCharges}, we know that for $v\in \bar{B}$, we have $\sum_{u\in T_v}\chrg{u}{v}\leq\frac{w(v)}{2}$. Corollary~\ref{CorBoundwAstar} and Lemma~\ref{LemWeightsNeighborhoodsdminus1} from the analysis of SquareImp, combined with $w(\bar{B})\leq (1-\delta) \cdot w(A)$ and hence $w(A)-w(\bar{B})\geq \delta\cdot w(A)$ as well as the definition of $T_v$ for $v\in A$ lead to \begin{align*}
 	w(A^*)&\leq \frac{d-1}{2}\cdot w(A)+\sum_{u\in A^*:\chrg{u}{n(u)}>0} \chrg{u}{n(u)}\\
 	&= \frac{d-1}{2}\cdot w(A)+\sum_{v\in A}\sum_{u\in T_v}\chrg{u}{v}\\
 	&\leq \frac{d-1}{2}\cdot w(A)+\sum_{v\in \bar{B}}\frac{w(v)}{2}+\sum_{v\in A\backslash \bar{B}}\frac{1-\epsilon}{2}\cdot w(v)\\
 	&= \frac{d}{2}\cdot w(A)-\frac{\epsilon}{2}\cdot (w(A)-w(\bar{B}))\\
 	&\leq \frac{d-\epsilon\delta}{2}\cdot w(A),
 	\end{align*} proving the assertion.
 \end{proof}
 \begin{proposition}
  $B\rightarrow B^*,v\mapsto t(v)$ is a bijection with inverse map $n\upharpoonright B^*$.\label{Proptbijection}
 \end{proposition}
\begin{proof}
Surjectivity follows from the definition of $B^*$, injectivity from the facts that each $u\in A^*$ may send positive charges to at most one $v\in A$ and that we have $t(v)\in T_v$ for all $v\in B$ by definition. As for $u\in B^*$, $n(u)$ is the unique vertex in $A$ that $u$ can send positive charges to, we must have $u=t(n(u))$, which implies the second part of the assertion.
\end{proof}
 \begin{proof}[Proof of Lemma~\ref{SmallRemainingNeighbourhood}]
  	Let $v\in B$ and $u:=t(v)$. By the definition of $u=t(v)$, we have $v\in N(u,A)$, $v=n(u)$ and $u$ is single. This yields
  	\begin{align*}
 	 w(N(u,A)\backslash\{v\})=  w(N(u,A))-w(v)\leq (1+\sqrt{\epsilon})\cdot w(v)-w(v)=\sqrt{\epsilon}\cdot w(v).
 	\end{align*}
  \end{proof}
   \begin{proof}[Proof of Lemma~\ref{LemBoundWsqV}]
   If $w(v)\leq w(t(v))$ this is clear since all weights are positive and $\epsilon >0$ by \eqref{constants5}. Therefore, assume that $w(t(v))<w(v)$. By the definition of single vertices, we obtain \[w(v)\leq \frac{1}{1-\sqrt{\epsilon}}\cdot w(t(v))=\left(1+\frac{\sqrt{\epsilon}}{1-\sqrt{\epsilon}}\right)\cdot w(t(v))\leq (1+2\sqrt{\epsilon})\cdot w(t(v)) \]since $0\leq \epsilon<\frac{1}{4}$ by \eqref{constants5}.
   Consequently, our assumption $w(t(v))<w(v)$ and the fact that all weights are positive yield
   \begin{align*}w^2(v)&\leq (1+2\sqrt{\epsilon})^2\cdot w^2(t(v))=(1+4\sqrt{\epsilon}+4\epsilon)\cdot w^2(t(v))\\&\leq w^2(t(v))+(4\sqrt{\epsilon}+4\epsilon)\cdot w^2(v)\end{align*} as claimed.
  \end{proof}
  \begin{proof}[Proof of Lemma~\ref{Amaximal}]
 	Let $u\in A'^*$ with $N(u,A')=\emptyset$ and define $T:=\{t(v), v\in N(u,B)\}$. We show that $T\cup\{u\}$ yields a local improvement of $w^2(A)$. First, as $B\subseteq A$, Proposition~\ref{PropSizeBound} and Proposition~\ref{Proptbijection} tell us that $|T|=|N(u,B)|\leq d-1, $ so $T\cup\{u\}$ contains at most $d\leq (d-1)^2+d-1$ vertices since $d\geq 3$. The neighbors of $T\cup\{u\}$ in $A$ can be split into the neighbors $N(u,B)$ of $u$ in $B$ and the neighbors of $T$ in $A$ that are not contained in $N(u,B)$, because $N(u,A\backslash B)=N(u,A')=\emptyset$ by choice of $u$ (see Figure~\ref{FigImprovementIfNoNeighbours}). 
 	 \begin{figure}[t]
 	 \centering
   \begin{tikzpicture}[scale = 0.7, mynode/.style = {circle, draw = black, thick, fill = black, inner sep = 0mm, minimum size = 2mm}] 
    \draw[draw, rounded corners] (0,0) rectangle (6,2);
    \node at (0.5,0.5) {$A'^*$};
   \draw[draw, rounded corners] (0,5) rectangle (6,7);
   \node at (0.5,6.5) {$A'$};
   \draw[draw, rounded corners] (7,0) rectangle (13,2);
   \node at (7.5,0.5) {$B^*$};
   \draw[draw, rounded corners] (7,5) rectangle (13,7);
   \node at (7.5,6.5) {$B$};
   \draw[draw, rounded corners] (14,0) rectangle (17,2);
   \node at (14.5,0.5) {$P$};
   \node at (-1,1) {$A^*$};
   \node at (-1,6) {$A$};
   \foreach \x in {8,12}
   {
      \node[mynode] (L) at (\x,1){};
      \node[mynode] (U) at (\x,6){};
      \draw[very thick] (L)--(U);
   }
   \foreach \x in {9,10,11}
   {
      \node (L) at (\x,1){};
      \node(U) at (\x,6){};
      \draw[very thick, blue] (L)--(U);
      \node[mynode] at (\x,1){};
      \node[mynode] at (\x,6){};
   }
   \foreach \x in {1,2,3,4,5}
   {
      \node[mynode] (L) at (\x,1){};
      \node[mynode] (U) at (\x,6){};
   }
   \foreach \x in {15,16}
   {
      \node[mynode] (L) at (\x,1){};
   }
   \draw[dashed] (8,1)--(1,6);
   \draw[dashed, orange] (10,1)--(4,6);
   \draw[dashed, orange] (10,1)--(2,6);
   \draw[dashed] (12,1)--(5,6);
   \draw[dashed, orange] (9,1)--(12,6);
  \draw (10,6)--(3,1);
   \draw (11,6)--(16,1);
   \draw (9,6)--(1,1);
   \draw (8,6)--(2,1);
   \draw(2,6)--(4,1);
   \draw(1,6)--(4,1);
   \draw (12,6)--(15,1);
   \draw[blue] (5,1)--(9,6);
   \draw[blue] (5,1)--(10,6);
   \draw[blue] (5,1)--(11,6);
   \draw (2,1)--(3,6);
   \draw (3,1)--(4,6);
   \draw (1,1)--(5,6);
   \node[mynode, draw = green!70!black, fill = green!70!black] at (5,1){};
   \node[mynode, draw = red!70!black, fill = red!70!black] at (9,6){};
   \node[mynode, draw = red!70!black, fill = red!70!black] at (10,6){};
   \node[mynode, draw = red!70!black, fill = red!70!black] at (11,6){};
   \draw[rounded corners, draw = red!70!black] (8.7,5.7) rectangle (11.3,6.3);
   \node at (9,7.5) {\textcolor{red!70!black}{$N(u,B)$}};
   \node[mynode, draw = green!70!black, fill = green!70!black] at (9,1){};
   \node[mynode, draw = green!70!black, fill = green!70!black] at (10,1){};
   \node[mynode, draw = green!70!black, fill = green!70!black] at (11,1){};
   \draw[rounded corners, draw = green!70!black] (8.7,0.7) rectangle (11.3,1.3);
   \node[mynode, draw = orange, fill = orange] at (2,6){};
   \node[mynode, draw = orange, fill = orange] at (4,6){};
   \node[mynode, draw = orange, fill = orange] at (12,6){};
   \node at (12.5,0.5) {\textcolor{green!70!black}{$T$}};
   \node at (5.5,0.5){\textcolor{green!70!black}{$u$}}; 
   \node at (3,7.5) {\textcolor{orange}{$N(T,A)\backslash N(u,B)$}};
   \node at (16,6) {\textcolor{red!50!blue}{$N(T\cup\{u\},A)$}};
   \draw[draw = red!50!blue, rounded corners] (1.5,5.5) rectangle (2.5,6.5);
   \draw[draw = red!50!blue, rounded corners] (3.5,5.5) rectangle (4.5,6.5);
   \draw[draw = red!50!blue, rounded corners] (8.5,5.5) rectangle (12.5,6.5);
   \end{tikzpicture}
 \caption{The situation in Lemma~\ref{Amaximal}. Dashed lines indicate edges from vertices in $B^*$ to vertices in $A$ of significantly lower weight, thick vertical lines mark the edges connecting $v\in B$ to $t(v)\in B^*$.}\label{FigImprovementIfNoNeighbours}
\end{figure}
Hence, we get
 	\begin{equation}w^2(N(T\cup\{u\},A))\leq w^2(N(u,B))+w^2(N(T,A)\backslash N(u,B)).\label{BoundWSNeighbourhood}\end{equation}
 	By Lemma~\ref{LemBoundWsqV} and Proposition~\ref{Proptbijection}, we know that \begin{align}
 	                                           w^2(N(u,B))=&\sum_{v\in N(u,B)}w^2(v)\leq \sum_{v\in N(u,B)} w^2(t(v))+(4\sqrt{\epsilon}+4\epsilon)\cdot w^2(v)\notag\\=&w^2(T)+(4\sqrt{\epsilon}+4\epsilon)\cdot w^2(N(u,B)).\label{Boundw^2(N(u,B))}
 	                                          \end{align}

 	Next, Lemma~\ref{SmallRemainingNeighbourhood} and Proposition~\ref{Proptbijection} tell us that
 	\begin{align} w^2(N(T,A)\backslash N(u,B)) &\leq \sum_{t\in T} w^2(N(t,A)\backslash N(u,B))\notag\\ 
 	 &= \sum_{v\in N(u,B)} w^2(N(t(v),A)\backslash N(u,B))\notag\\
 	 &\leq\sum_{v\in N(u,B)} w^2(N(t(v),A)\backslash\{v\})\notag\\
 	 &\leq \sum_{v\in N(u,B)} \epsilon\cdot w^2(v)\notag\\
 	 &=\epsilon\cdot w^2(N(u,B)).\label{Boundw^2(N(T,A)backslashN(u,B))}
 	\end{align} Combining \eqref{BoundWSNeighbourhood}, \eqref{Boundw^2(N(u,B))} and \eqref{Boundw^2(N(T,A)backslashN(u,B))}, we obtain
 	\[w^2(N(T\cup\{u\},A))\leq (4\sqrt{\epsilon}+5\epsilon)\cdot w^2(N(u,B)) +w^2(T).\]
 	As $u\in A'^*=A^*\backslash(B^*\cup P)$ and by definition of $P$, we know that \[w(N(u,B))\leq w(N(u,A))\leq 3\cdot w(u), \]so \[(4\sqrt{\epsilon}+5\epsilon)\cdot w^2(N(u,B))\leq 9\cdot(4\sqrt{\epsilon}+5\epsilon)\cdot w^2(u)<w^2(u) \]by \eqref{constants2} and since $w(u)>0$.
 	Consequently, \[w^2(N(T\cup\{u\},A))< w^2(u)+w^2(T)=w^2(T\cup\{u\}) \] since $u\in A'^*=A^*\backslash(B^*\cup P)$ and $T\subseteq B^*$ and we have found a local improvement as claimed.
 \end{proof}
 \begin{proof}[Proof of Lemma~\ref{LemImprovedApproxIfSmallCharges}]
Observing that $G[A'\cup A'^*]$ is $d$-claw free as an induced subgraph of $G$, Corollary~\ref{CorBoundwAstar} and Lemma~\ref{LemWeightsNeighborhoodsdminus1} tell us that
 \begin{align*} w(A'^*) &\leq \sum_{u\in A'^*} \frac{w(N(u,A'))}{2}+\sum_{u\in A'^*:\chrgp{u}{n'(u)}>0}\chrgp{u}{n'(u)}\\
  &\leq \frac{d-1}{2}\cdot w(A')+\sum_{v\in A'}\sum_{u\in T'_v} \chrgp{u}{v}\\
  &\leq \frac{d-1}{2}\cdot w(A')+\sum_{v\in A'} \frac{d+2}{4}\cdot w(v)\\
  &= \frac{d-1}{2}\cdot w(A')+ \frac{d+2}{4}\cdot w(A')\\
  &=\frac{3d}{4} \cdot w(A').
 \end{align*}
Moreover, by Lemma~\ref{NeighbourhoodLargeCharges} and by definition of $t(v)$ for $v\in B$, we have \[w(B^*)=w(\{t(v):v\in B\})\leq (1+\sqrt{\epsilon})\cdot w(B). \] By assumption, we further know that $w(P)\leq \epsilon\delta\cdot w(A)$ as well as $w(B)\geq (1-\delta-\frac{25}{12}\cdot\epsilon\delta)\cdot w(A)$ and $w(A')=w(A)-w(B)$. Putting everything together, we obtain
 \begin{align*}
 w(A^*)&=w(B^*)+w(A'^*)+w(P)&\\
 &\leq (1+\sqrt{\epsilon})\cdot w(B)+\frac{3d}{4}\cdot (w(A)-w(B))+\epsilon\delta\cdot w(A)&\\
 &= \left(\frac{3d}{4}+\epsilon\delta\right)\cdot w(A)-\left(\frac{3d}{4}-(1+\sqrt{\epsilon})\right)\cdot w(B)&\text{ $|$ \eqref{constants7}}\\
 &\leq  \left(\frac{3d}{4}+\epsilon\delta\right)\cdot w(A)-\left(\frac{3d}{4}-(1+\sqrt{\epsilon})\right)\cdot \left(1-\delta-\frac{25}{12}\cdot\epsilon\delta \right)\cdot w(A)&\\
 &=\left((1+\sqrt{\epsilon})\cdot \left(1-\delta-\frac{25}{12}\cdot\epsilon\delta \right)+\frac{3d}{4}\cdot \left(\delta+\frac{25}{12}\cdot\epsilon\delta\right) +\epsilon\delta\right)\cdot w(A)&\text{ $|$ \eqref{constants3}}\\
 &\leq \frac{d-\epsilon\delta}{2}\cdot w(A),
 \end{align*} which concludes the proof.
\end{proof}
\begin{proof}[Proof of Lemma~\ref{LemGetImprovementOldInstance}]
 Assume that the assertion does not hold and pick $v_0\in A'$ such that \[\sum_{u\in T'_{v_0}}\chrgp{u}{v_0}> \frac{d+2}{4}\cdot w(v_0).\]
 Let $R:=\{t(v):v\in N(T'_{v_0},B)\}$. We show that $T'_{v_0}\cup R$ yields a local improvement of $w^2(A)$, contradicting the termination criterion of our algorithm.\\
 	As $T'_{v_0}\subseteq N(v_0,A^*)$, Proposition~\ref{PropSizeBound} implies that $|T'_{v_0}|\leq d-1$. Given that for $u\in T'_{v_0}\subseteq A^*$, $N(u,B)\subseteq N(u,A)$ can contain at most $d-1$ elements by Proposition~\ref{PropSizeBound}, Proposition~\ref{Proptbijection} implies that $|R|=|N(T'_{v_0},B)|\leq (d-1)^2$. Hence, the total size of our improvement is at most $(d-1)^2+(d-1)$.\\
 	As $\chrgp{u}{v_0}>0$ for all $u\in T'_{v_0}$, Lemma~\ref{LemPropPositiveCharges} shows that \[w^2(u)-w^2(N(u,A')\backslash\{v_0\})\geq 2\cdot\chrgp{u}{v_0}\cdot w(v_0) \]for all $u\in T'_{v_0}$. \\
 	Additionally, for $u\in T'_{v_0}$ with $w(u)\geq 4\cdot w(v_0)$, we get \[2\cdot w(u)-w(N(u,A'))=2\cdot\chrgp{u}{v_0} \]and therefore \[w(N(u,A'))=2\cdot w(u)-2\cdot \chrgp{u}{v_0}. \] As $v_0$ is the heaviest neighbor of $u$ in $A'$ by definition of charges, we further obtain \begin{align*}&w^2(N(u,A')\backslash\{v_0\})\leq w^2(N(u,A'))\leq \sum_{v\in N(u,A')}w(v)\cdot w(v_0)\\&=w(N(u,A'))\cdot w(v_0) = (2\cdot w(u)-2\cdot\chrgp{u}{v_0})\cdot w(v_0)\\
 	&\leq 2\cdot w(u)\cdot\frac{w(u)}{4}-2\cdot\chrgp{u}{v_0}\cdot w(v_0) = \frac{w(u)^2}{2}-2\cdot\chrgp{u}{v_0}\cdot w(v_0).\end{align*}
 	As a consequence, \[\frac{w(u)^2}{2}-w^2(N(u,A'\backslash\{v_0\}))\geq 2\cdot\chrgp{u}{v_0}\cdot w(v_0).\]
 	Let $S'_{v_0}:=\{u\in T'_{v_0}:w(u)\geq 4\cdot w(v_0)\}$.
 	Then \[\sum_{u\in T'_{v_0}}\chrgp{u}{v_0}> \frac{d+2}{4}\cdot w(v_0),\] together with the previous considerations and $w(v_0)>0$, implies that
 	\begin{align*} &\sum_{u\in T'_{v_0}}w^2(u)-w^2(N(u,A')\backslash\{v_0\}) \\= & \sum_{u\in S'_{v_0}}\frac{w^2(u)}{2}-w^2(N(u,A')\backslash\{v_0\})+\sum_{u\in T'_{v_0}\backslash S'_{v_0}}w^2(u)-w^2(N(u,A')\backslash\{v_0\})\\&+\sum_{u\in S'_{v_0}} \frac{w^2(u)}{2}\\
 	 \geq & \sum_{u\in S'_{v_0}}  2\cdot\chrgp{u}{v_0}\cdot w(v_0)+ \sum_{u\in T'_{v_0}\backslash S'_{v_0}}2\cdot\chrgp{u}{v_0}\cdot w(v_0)\\&+\sum_{u\in S'_{v_0}} \frac{w^2(u)}{2}\\
 	 &=\sum_{u\in T'_{v_0}}  2\cdot\chrgp{u}{v_0}\cdot w(v_0)+\sum_{u\in S'_{v_0}} \frac{w^2(u)}{2}\\
 	 &> \left(1+\frac{d}{2}\right)\cdot w^2(v_0)+\sum_{u\in S'_{v_0}} \frac{w^2(u)}{2}.
 	\end{align*}
This implies 
 	\[\sum_{u\in T'_{v_0}}w^2(u)>w^2(v_0)+\sum_{u\in T'_{v_0}}w^2(N(u,A')\backslash\{v_0\})+\sum_{u\in S'_{v_0}}\frac{w^2(u)}{2}+\frac{d}{2}\cdot w^2(v_0) \] and hence
 	\begin{align}w^2(T'_{v_0}) &> w^2(N(T'_{v_0},A')) +\sum_{u\in S'_{v_0}}\frac{w^2(u)}{2}+ \frac{d}{2}\cdot w^2(v_0)\notag\\
 	&\geq w^2(N(T'_{v_0},A'))+\sum_{u\in S'_{v_0}}\frac{w^2(u)}{2} + \sum_{u\in T'_{v_0}\backslash S'_{v_0}}\frac{w^2(u)}{32}\notag\\
 	&\geq w^2(N(T'_{v_0},A'))+\sum_{u\in T'_{v_0}}\frac{w^2(u)}{32}\notag\\
 	&= w^2(N(T'_{v_0},A'))+\frac{1}{32}\cdot w^2(T'_{v_0})\label{EqlargeEnoughSlack}
 	\end{align}
         since  $|T'_{v_0}|\leq d-1$ and $w(u)\leq 4\cdot w(v_0)$ for $u\in T'_{v_0}\backslash S'_{v_0}$.
 	We know that we can split the neighbors of $T'_{v_0}\cup R$ in $A$ into the neighbors $N(T'_{v_0},A')$ of $T'_{v_0}$ in $A'$, the neighbors $N(T'_{v_0},B)$ of $T'_{v_0}$ in $B$ and the neighbors of $R$ that we did not consider yet, i.e.\ $N(R, A)\backslash N(T'_{v_0},A)$ (see Figure~\ref{FigImprovementOldInstance}).
 	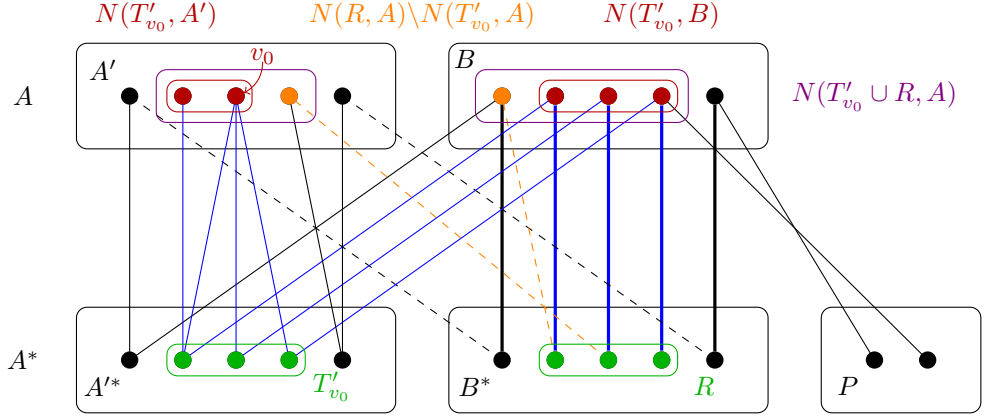
\begin{figure}[t]
 	\centering
  \begin{tikzpicture}[scale = 0.7, mynode/.style = {circle, draw = black, thick, fill = black, inner sep = 0mm, minimum size = 2mm}] 
   \draw[draw, rounded corners] (0,0) rectangle (6,2);
   \node at (0.5,0.5) {$A'^*$};
   \draw[draw, rounded corners] (0,5) rectangle (6,7);
   \node at (0.5,6.5) {$A'$};
   \draw[draw, rounded corners] (7,0) rectangle (13,2);
   \node at (7.5,0.5) {$B^*$};
   \draw[draw, rounded corners] (7,5) rectangle (13,7);
   \node at (7.3,6.7) {$B$};
   \draw[draw, rounded corners] (14,0) rectangle (17,2);
   \node at (14.5,0.5) {$P$};
   \node at (-1,1) {$A^*$};
   \node at (-1,6) {$A$};
   \foreach \x in {8,12}
   {
      \node[mynode] (L) at (\x,1){};
      \node[mynode] (U) at (\x,6){};
      \draw[very thick] (L)--(U);
   }
    \foreach \x in {9,10,11}
   {
    \node (L) at (\x,1){};
    \node(U) at (\x,6){};
    \draw[very thick, blue] (L)--(U);
      \node[mynode] at (\x,1){};
      \node[mynode] at (\x,6){};
   }
   \foreach \x in {1,2,3,4,5}
   {
      \node[mynode] (L) at (\x,1){};
      \node[mynode] (U) at (\x,6){};
   }
   \foreach \x in {15,16}
   {
      \node[mynode] (L) at (\x,1){};
   }
   \draw[dashed] (8,1)--(1,6);
   \draw[dashed, orange] (10,1)--(4,6);
   \draw[dashed] (12,1)--(5,6);
   \draw[dashed, orange] (9,1)--(8,6);
   \draw[blue] (10,6)--(3,1);
   \draw (11,6)--(16,1);
   \draw (8,6)--(1,1);
   \draw[blue] (9,6)--(2,1);
   \draw (12,6)--(15,1);
   \draw[blue] (3,6)--(2,1);
   \draw[blue] (3,6)--(3,1);
   \draw[blue] (3,6)--(4,1);
   \draw[blue] (4,1)--(11,6);
   \draw (5,1)--(5,6);
   \draw (5,1)--(4,6);
   \draw (1,1)--(1,6);
   \draw[blue] (2,1)--(2,6);
   \node[mynode, draw = red!70!black, fill = red!70!black] (V0) at (3,6){};
  \node (V0Label) at (3.5,6.75) {\textcolor{red!70!black}{$v_0$}};
   \draw[red!70!black, ->] (3.5,6.65) to [bend left = 30] (V0);
   \node[mynode, draw = red!70!black, fill = red!70!black]  at (2,6){};
   \node[mynode, draw = red!70!black, fill = red!70!black] at (9,6){};
   \node[mynode, draw = red!70!black, fill = red!70!black] at (10,6){};
   \node[mynode, draw = red!70!black, fill = red!70!black] at (11,6){};
   \draw[rounded corners, draw = red!70!black] (1.7,5.7) rectangle (3.3,6.3);
   \draw[rounded corners, draw = red!70!black] (8.7,5.7) rectangle (11.3,6.3);
   \node at (11,7.5) {\textcolor{red!70!black}{$N(T'_{v_0},B)$}};
   \node at (1.5,7.5) {\textcolor{red!70!black}{$N(T'_{v_0},A')$}};
   \node[mynode, draw = green!70!black, fill = green!70!black] at (9,1){};
   \node[mynode, draw = green!70!black, fill = green!70!black] at (10,1){};
   \node[mynode, draw = green!70!black, fill = green!70!black] at (11,1){};
   \node[mynode, draw = green!70!black, fill = green!70!black] at (2,1){};
   \node[mynode, draw = green!70!black, fill = green!70!black] at (3,1){};
   \node[mynode, draw = green!70!black, fill = green!70!black] at (4,1){};
   \draw[rounded corners, draw = green!70!black] (8.7,0.7) rectangle (11.3,1.3);
   \draw[rounded corners, draw = green!70!black] (1.7,0.7) rectangle (4.3,1.3);
   \node at (11.8,0.5) {\textcolor{green!70!black}{$R$}};
    \node at (4.8,0.5){\textcolor{green!70!black}{$T'_{v_0}$}};
   \node[mynode, draw = orange, fill = orange] at (8,6){};
   \node[mynode, draw = orange, fill = orange] at (4,6){};
   \draw[rounded corners, draw = red!50!blue] (1.5,5.5) rectangle (4.5,6.5);
   \draw[rounded corners, draw = red!50!blue] (7.5,5.5) rectangle (11.5,6.5);
   \node at (6.5,7.5){\textcolor{orange}{$N(R,A)\backslash N(T'_{v_0},A)$}};
   \node at (15,6){\textcolor{red!50!blue}{$N(T'_{v_0}\cup R,A)$}};
   
  \end{tikzpicture}
\caption{The situation in Lemma~\ref{LemGetImprovementOldInstance}. Dashed lines indicate edges from vertices in $B^*$ to vertices in $A$ of significantly lower weight, thick vertical lines mark the edges connecting $v\in B$ to $t(v)\in B^*$.}\label{FigImprovementOldInstance}
 \end{figure}
 	For $u\in R$ and $v:=n(u)\in N(T'_{v_0},B)\subseteq N(T'_{v_0},A)$, we have $u=t(v)$ by Proposition~\ref{Proptbijection} and  $w(N(u,A)\backslash\{v\})\leq \sqrt{\epsilon}\cdot w(v)$ by Lemma~\ref{SmallRemainingNeighbourhood}. This shows that \[w^2(N(R,A)\backslash N(T'_{v_0},A))\leq \epsilon\cdot w^2(N(T'_{v_0},B)). \] As $T'_{v_0}\subseteq A'^*=A^*\backslash(B^*\cup P)$, we have \[w^2(N(u,B))\leq w^2(N(u,A))\leq 9\cdot w^2(u) \] for all $u\in T'_{v_0}$, showing that \[w^2(N(T'_{v_0},B))\leq w^2(N(T'_{v_0},A))\leq \sum_{u\in T'_{v_0}}w^2(N(u,A))\leq 9\sum_{u\in T'_{v_0}} w^2(u)=9\cdot w^2(T'_{v_0}) \] and hence
 	\begin{equation} w^2(N(R,A)\backslash N(T'_{v_0},A))\leq \epsilon\cdot w^2(N(T'_{v_0},B))\leq 9\epsilon\cdot w^2(T'_{v_0}).\label{lowweightneighborsofR}\end{equation}
 	Finally, Lemma~\ref{LemBoundWsqV} and Proposition~\ref{Proptbijection} yield \begin{align}w^2(N(T'_{v_0},B))&\leq w^2(R)+(4\sqrt{\epsilon}+4\epsilon)\cdot w^2(N(T'_{v_0},B))\notag\\
 	&\leq w^2(R)+(4\sqrt{\epsilon}+4\epsilon)\cdot 9\cdot w^2(T'_{v_0})\notag\\
 	&= w^2(R)+(36\sqrt{\epsilon}+36\epsilon)\cdot w^2(T'_{v_0})\label{EqBoundWeightneighborsTprime}.\end{align} Combining \eqref{EqlargeEnoughSlack}, \eqref{lowweightneighborsofR} and \eqref{EqBoundWeightneighborsTprime}, we get
 	\begin{align*}
 	w^2(N(T'_{v_0}\cup R,A))=&\quad w^2(N(T'_{v_0},A'))+w^2(N(T'_{v_0},B))\\
 	&+w^2(N(R,A)\backslash N(T'_{v_0},A)) \\
 	<&\quad w^2(T'_{v_0})-\frac{1}{32}\cdot w^2(T'_{v_0})+w^2(R)\\
 	&+(36\sqrt{\epsilon}+45\epsilon)\cdot w^2(T'_{v_0})\\
 	\leq&\quad w^2(T'_{v_0})+w^2(R)-\left(\frac{1}{32}-(36\sqrt{\epsilon}+45\epsilon)\right)w^2(T'_{v_0})\\
 	\leq &\quad w^2(T'_{v_0})+w^2(R)\\
 	=&\quad w^2(T'_{v_0}\cup R)
 	\end{align*} by \eqref{constants4} and since $T'_{v_0}\subseteq A'^*$ and $R\subseteq B^*$ are disjoint. So we indeed get a local improvement of $w^2(A)$, a contradiction.
 	
 \end{proof}
\end{document}